\documentclass[useAMS,usenatbib,usegraphicx]{mn2e}

\usepackage{amssymb} 
\usepackage{textcomp} 
\usepackage{appendix}
\usepackage{amsmath} 
\usepackage{fixltx2e} 
\usepackage{multirow} 
\usepackage{enumerate} 
\usepackage{mathptmx} 
\usepackage{lscape}  
\usepackage{hyperref} 

\newcommand{\kms}{\rm km\ s^{-1}}
\newcommand{\ergs}{\rm erg\ s^{-1}}
\newcommand{\flux}{\rm erg\ s^{-1}\ cm^{-2}}
\newcommand{\mic}{\mbox{$\mu$m}}

\newcommand{\kev}{\rm keV}

\let\AAold\AA
\renewcommand{\AA}{\text{\AAold}}

\newcommand{\lbol}{L_{\rm bol}}
\newcommand{\lledd}{L / L_{\rm{Edd}}}
\newcommand{\mbh}{M_{\rm BH}}

\newcommand{\nln}{\nu L_{\nu}}

\newcommand{\ebv}{E_{\rm B-V}}
\newcommand{\hc}{h_3}
\newcommand{\hd}{h_4}

\newcommand{\msun}{{\rm M_{\odot}}}

\newcommand{\hi}{\text{H~{\sc i}}}
\newcommand{\hii}{\text{H~{\sc ii}}}
\newcommand{\nii}{\text{[N~{\sc ii}]}}
\newcommand{\sii}{\text{[S~{\sc ii}]}}

\newcommand{\oi}{\text{[O~{\sc i}]}}
\newcommand{\oiii}{\text{[O~{\sc iii}]}}

\newcommand{\feii}{\text{[Fe~{\sc ii}]}}

\newcommand{\Ha}{\text{H$\alpha$}}

\newcommand{\Hb}{\text{H$\beta$}}

\newcommand{\aap}{A\&A}
\newcommand{\apjl}{ApJ}
\newcommand{\apjs}{ApJS}
\newcommand{\apj}{ApJ}
\newcommand{\aj}{AJ}
\newcommand{\mnras}{MNRAS}
\newcommand{\lbha}{L_{\rmn{bH\alpha}}}
\newcommand{\lbhb}{L_{\rmn{bH\beta}}}
\newcommand{\fbha}{F_{\rmn{bH\alpha}}}
\newcommand{\lnha}{L_{\rmn{nH\alpha}}}
\newcommand{\lnhb}{L_{\rmn{nH\beta}}}
\newcommand{\fnha}{F_{\rmn{nH\alpha}}}
\newcommand{\lha}{L_{\rmn{H\alpha}}}
\newcommand{\loiii}{L_{\oiii}}
\newcommand{\foiii}{F_{\oiii}}

\newcommand{\lx}{L_{\rm X}}
\newcommand{\luv}{L_{\rm UV}}

\newcommand{\dv}{\Delta {\rm v}}
\newcommand{\dvt}{\Delta {\rm v}_{\rm total}}

\newcommand{\lagn}{L_{\rm AGN}}
\newcommand{\lhost}{L_{\rm host}}

\newcommand{\cfnlr}{{\rm CF}_{\rm NLR}}
\newcommand{\cfblr}{{\rm CF}_{\rm BLR}}

\renewcommand{\th}{$^{\rm th}$}

\newcommand{\smpsz}{3\,410}

\title[Type 1 low $z$ AGN. II. NLR relative luminosity]{Type 1 AGN at low $z$. II. The relative strength of narrow lines and the nature of intermediate type AGN} 
\author[Jonathan Stern and Ari Laor]
{Jonathan Stern\thanks{E-mail: \href{mailto:stern@physics.technion.ac.il}{stern@physics.technion.ac.il} (JS);\newline \href{mailto:laor@physics.technion.ac.il}{laor@physics.technion.ac.il} (AL)}
and Ari Laor\footnotemark[1]\\
Department of Physics, Technion -- Israel Institute of Technology, Haifa 32000, Israel}

\begin{document}

\maketitle

\begin{abstract}
We explore the relative strength of the narrow emission lines in an SDSS based sample of broad \Ha\ selected AGN, defined
in paper I. We find a decrease in the narrow to broad \Ha\ luminosity ($\lbha$) ratio with increasing $\lbha$, such that both $L(\oiii\ \lambda5007)$ and $L({\rm narrow\ \Ha})$ scale as $\propto \lbha^{0.7}$ for $10^{40}<\lbha<10^{45}\ \ergs$.
Following our earlier result that $\lbha \propto \lbol$, this trend indicates that the relative narrow line luminosity 
decreases with increasing $\lbol$. We derive $\lbol/10^{43}\ \ergs = 4000 ( L (\oiii)/10^{43}\ \ergs)^{1.39}$.
This implies that the bolometric correction factor, $\lbol/L(\oiii)$, decreases from $3\,000$ at $\lbol=10^{46.1}\ \ergs$ to $300$ at $\lbol=10^{42.5}\ \ergs$. At low luminosity, the narrow component dominates the observed \Ha\ profile, and most type 1 AGN appear as intermediate type AGN. Partial obscuration or extinction cannot explain the dominance of intermediate type AGN at low luminosity, and the most likely mechanism is a decrease in the narrow line region covering factor with increasing $\lbol$. 
Deviations from the above trend occur in objects with $\lledd \lesssim 10^{-2.6}$, probably due to the transition to LINERs with
suppressed \oiii\ emission, and in objects with $\mbh > 10^{8.5}\ \msun$, probably due to the dominance of radio loud AGN, and associated enhanced \oiii\ emission.
\end{abstract}


\section{Introduction}

The circumnuclear gas located on 1 -- 1\,000 pc scale from active galactic nuclei (AGN) is an important constituent in several intriguing, but poorly known processes. It is the source of AGN fuel, and it absorbs AGN energy and momentum via winds, radiation and/or jets. Understanding the properties of this gas is crucial to constrain theories of these processes, and their possible connection to star formation in the host bulge, which is implied from the central black hole -- bulge relations (Magorrian et al. 1998, Ferrarese \& Merritt 2000). 

Some of the properties of this gas have been inferred by examining how photons from the central source are reprocessed into secondary radiation. The AGN radiation is mainly converted either into thermal IR emission from dust grains embedded in the gas, or to permitted and forbidden emission lines. These emission lines are observed with widths of $\sim300\ \kms$, typical of the galaxy potential. The region that emits the lines is termed the narrow line region (NLR), as opposed to the broad line region (BLR, $\sim3\,000\ \kms$), which is located on smaller scales, within the black hole gravitational sphere of influence. The NLR is used to constrain the radial distribution of the circumnuclear gas (Ferguson et al. 1997), and to infer the AGN bolometric luminosity $\lbol$ when other measures are not available (e.g. Kauffmann \& Heckman 2009; Schawinski et al. 2010). This circumnuclear gas may also block our line of sight to the inner ionizing source and BLR, 
thus playing a role in the unobscured (type 1) / obscured (type 2) AGN dichotomy. 

How does the distribution of the circumnuclear gas depend on $\lbol$ and black hole mass $\mbh$? There is some evidence that the covering factor CF of the IR emitting gas decreases with $\lbol$ (Maiolino et al. 2007, Treister et al. 2008), as does the type 2 / type 1 ratio (Hasinger 2008). The dependence of the NLR covering factor, $\cfnlr$, on luminosity has not been directly explored, though there may be some indications that $\cfnlr$ is higher at low AGN luminosity (Ludwig et al. 2012). Also, at high $\lbol$, where the host contribution to the continuum is negligible, a decrease in \oiii\ equivalent width with luminosity has been observed (Sulentic et al. 2004, Netzer et al. 2006, Ludwig et al. 2009), which may also imply a luminosity-dependent $\cfnlr$. 

Moreover, in the classical type 1 / type 2 AGN division, the broad lines either dominate the permitted lines or are not observed at all. This division is unsatisfactory, as the broad and narrow components of the permitted lines show a large range in flux ratio between different objects (e.g. fig. 1 of Stern \& Laor 2012, hereafter Paper I). Following Osterbrock (1977), `intermediate types' were added to the scheme -- type 1.2s, 1.5s and 1.8s, with larger numbers corresponding to higher narrow to broad flux ratios.
If in type 1 AGN we have a direct view of the BLR, and in type 2 AGN the BLR is obscured, what then are the intermediate types?  
A simple interpolation of the above scheme may imply that the BLR is either partially obscured or undergoes extinction by optically thin dust. Partial obscuration is clearly a factor in AGN, as can be seen in spectropolarimetric surveys (e.g. Tran 2003) that show objects in which a `hidden' BLR is revealed in the polarized spectrum. 
Evidence for dust reddening of the BLR has also been found in intermediate type samples (e.g. Goodrich et al. 1995). 
However, Trippe et al. (2010) find that the majority of intermediate types in their sample are inconsistent with reddening of the BLR, which suggests another mechanism may be important in driving the intermediate type phenomenon. An alternative mechanism is that intermediate type AGN are not obscured, like `standard' type 1s, but have a large $\cfnlr/\cfblr$ ratio. As we show below, this mechanism appears to dominate in low luminosity AGN.

In this paper, we use the type 1 sample described in Paper I. The sample is based on the detection of broad \Ha\ emission, which
in some objects is relatively weak compared to the narrow \Ha\ emission, classifying them as intermediate type AGN. The
sample spans $10^{6}<\mbh<10^{9.5}\ \msun$ and $10^{42}<\lbol<10^{46}\ \ergs$. We derive the NLR luminosity and fraction of intermediate types, for different $\lbol$ and $\mbh$. 
Using these relations between the NLR and AGN properties, we explore various mechanisms proposed to explain the intermediate type phenomenon. 
We also calibrate the luminosity-dependent $\lbol/\loiii$ ratio, for use of future studies based on type 2 AGN.

The paper is organized as follows. In \S2, we give a summary of the sample selection and emission line measurements described in Paper I, emphasizing the measurements of \oiii\ and narrow \Ha, used here. In \S3 we present our main result, the increase in relative NLR luminosity and in the fraction of intermediate type AGN with decreasing luminosity. The relative NLR luminosity is also compared to $\lledd$ and $\mbh$. In \S4 we compare our results to earlier studies. The possible physical mechanisms behind the observed trend are examined in \S5. We discuss the implications in \S6, and provide conclusions in \S7.

Throughout the paper, we assume a FRW cosmology with $\Omega_{\rm m}$ = 0.3, $\Lambda$ = 0.7 and $H_0 = 70\ \kms$ Mpc$^{-1}$.

\section{The sample}

The construction of the broad \Ha\ selected sample used in this paper is described in Paper I. In \S2.1 we provide a brief summary, emphasizing the details most relevant to this work. In \S2.2, we discuss a change in the host subtraction procedure. In \S2.3 and \S2.4 we elaborate on the measurement of the \oiii\ luminosity $\loiii$ and the narrow \Ha\ luminosity $\lnha$, which are the basis of the analysis in this paper, and were not used previously.

\subsection{The T1 sample creation}

The type 1 (T1) sample is selected from the 7\th\ data release of the Sloan Digital Sky Survey (SDSS DR7; Abazajian et al. 2009). The SDSS obtained imaging of a quarter of the sky in five bands ({\it ugriz}; Fukugita et al. 1996) to a 95\% $r$ band completeness limit of 22.2 mag. Objects are selected for spectroscopy mainly due to their non-stellar colors (Richards et al. 2002), or extended morphology (Strauss et al. 2002). The spectrographs cover the wavelength range 3800\AA --9200\AA\ at a resolution of $\sim150\ \kms$, and are flux-calibrated by matching the spectra of simultaneously observed standard stars to their PSF magnitude (Adelman-McCarthy et al. 2008).

We use SDSS spectra which are classified as non-stellar and have $0.005<z<0.31$. To ensure a reliable decomposition of the broad and narrow components of \Ha, we use only spectra with S/N $>10$ and sufficient good spectral pixels in the vicinity of \Ha, as detailed in Paper I. These requirements are fulfilled by 232\,837 of the 1.6 million spectra in DR7. The spectra are corrected for foreground dust, using the maps of Schlegel et al. (1998) and the extinction law of Cardelli et al. (1989). We then subtract the host, as detailed below. Then, we subtract a featureless continuum, derived by interpolating the mean continuum level at 6125\AA--6250\AA\ and 6880\AA--7000\AA. The residual flux at 6250\AA--6880\AA\ ($\pm 14,000\ \kms$ from \Ha) is then summed, 
excluding regions $\pm 690\ \kms$ from the \oi\ $\lambda\lambda 6300,6363$, \nii\ $\lambda\lambda 6548,6583$, \sii\ $\lambda\lambda 6716,6730$ and \Ha\ narrow emission lines. We find 6\,986 objects with significant residual flux, which is potential broad \Ha\ emission. 

On the objects with significant residual near \Ha, we fit the profiles of the broad and narrow \Ha, \oiii\ $\lambda 5007$, and the \oi, \nii\ and \sii\ doublets mentioned above. Narrow lines are fit using 4\th -order Gauss-Hermite functions (GHs; van der Marel \& Franx 1993) and an up to 10\th -order GH is used for the broad \Ha\ profile. 
The following criteria are applied to the broad \Ha\ fit, in order to exclude objects in which the residual flux is not clearly BLR emission: the FWHM ($\dv$) of the fit is required to be in the range $1\,000 - 25\,000\ \kms$; the total flux of the fit, and its flux density at the line centre, are required to be significant. The T1 sample consists of the 3\,579 objects which passed these criteria. The broad \Ha\ luminosity ($\lbha$) and $\dv$ of the 3\,410 T1 objects used here (see below) are listed in Table 1. The selection effects implied by our selection criteria are detailed in Paper I.

We supplement the optical SDSS spectra in the T1 sample by photometric measurements in the UV and X-ray, as described below.

\subsubsection{$\luv$}

The GALEX mission (Martin et al. 2005, Morrissey et al. 2007) observed 2/3 of the sky in the FUV (effective wavelength 1528\AA) and NUV (2271\AA) bands. 
We search the GALEX GR6 catalog for objects within $5\arcsec$ of the T1 objects. Of the 89\% of the T1 objects observed by GALEX, 93\% have detections in both bands. We use UV fluxes derived from the 6\arcsec-radius `aperture 4', corrected for the PSF and for foreground dust assuming $A_{\rm{NUV}}/\ebv = 8.2$ and $A_{\rm{FUV}}/\ebv = 8.24$ (Wyder et al. 2007). In case of multiple detections per object, we co-add the observations, weighted by exposure time. For an object that was not detected, we assign an upper limit equal to the mean upper limit of the relevant GALEX survey (Morrissey et al. 2007), scaled to the exposure time at the field of the object. 
We derive $\luv \equiv \nln$(1528\AA), the luminosity of the shortest restframe wavelength available in all detected objects, by a power law interpolation between the two UV bands. The $\luv$ values are listed in Table 1.

\subsubsection{$\lx$}

The ROSAT All-Sky Survey (Voges et al. 1999, 2000) covers the entire celestial sphere in the $0.1 - 2.4$ \kev\ range, with positional uncertainties of $10\arcsec - 30\arcsec$. We find matches to 1\,561 (43\%) of the T1 sample within 50\arcsec\ of each T1 object, expecting 7 objects to have false matches. Only 0.4\% of the T1 objects were targeted for spectroscopy by the SDSS solely for being near a ROSAT source. Therefore, we do not expect our sample to be biased towards X-ray bright AGN. The ROSAT count rates are converted to $\lx \equiv \nln$(2~\kev) assuming a power-law X-ray spectrum with $\alpha_{\rm x}$ = 1.5 (Laor et al. 1994, Schartel et al. 1996), the $z$ of the optical match, and the Galactic $N_{\rm H}$ measurements of Stark et al. (1992). Upper limits in T1 objects without ROSAT detections are set to the typical limiting sensitivity of $10^{-12.5}\ \flux$. The $\lx$ values are listed in Table 1.

\subsection{Host subtraction}
In Paper I, we fit three galaxy eigenspectra (ESa) from Yip et al. 2004, and a $L_\lambda \propto \lambda^{-1.5}$ power law representing the AGN continuum, to wavelength regions devoid of strong emission lines (Paper I, appendix A2). We then account for the non-Balmer absorption lines near \Ha\ by subtracting the three fit ESa. However, since the Yip et al. ESa include emission lines which we interpolate over prior to the subtraction, the Balmer absorption features are not accounted for. This prescription sufficed for modeling the broad \Ha\ component, as the observed equivalent width (EW) in the T1 sample is typically $\sim100\AA$, and $>15\AA$ in 99\% of the objects. In contrast, the \Ha\ stellar absorption EW is typically 2\AA\ for old stars (see fig. 1 in Hao et al. 2005), and $<5\AA$ for a young stellar population (Groves et al. 2012). 

However, the apparent flux of the weaker narrow \Ha\ might be more strongly affected by stellar \Ha\ absorption. Therefore here, instead of a simple interpolation, we replace the emission lines in ES1 with absorption features of old stars from the Hao et al. (2005) ES1. The latter was derived from a principle component analysis of galaxies without emission lines. The fact that the continua in both the Yip et al. ES1 and the Hao et al. ES1 are dominated by old stars justifies the replacement. Technical details of this procedure are given in appendix A. 

Our fit does not account for the possible Balmer absorption from young stars. Their possible effect on $\lnha$ is addressed below.

We note that the \Ha\ absorption feature in Hao ES1 has $\sigma\sim 400\ \kms$, compared to the typical $\sigma\sim150\ \kms$ of other absorption features, indicating the intrinsic width dominates the width of the feature. Therefore, broadening the ES during the fit will have a negligible effect on the \Ha\ absorption feature in all but the most massive galaxies. Hence, we avoid this additional complexity in the fitting algorithm.

How does this adjustment of the Yip et al. ES1 affect the measured broad \Ha? In 98\% of the objects, the change in $\lbha$ and $\dv$, compared to the Paper I result, is $<0.1$ dex. This negligible change is expected due to the large broad \Ha\ EW. From inspection, 58 of the remaining 2\% (74 objects) have reasonable fits. The remaining 16 objects were not further processed due to their relatively small number, and they were removed from the sample. Four objects were fit with $\dv<1000\ \kms$, below our selection threshold, and therefore removed as well.

\subsection{The \oiii\ measurements}

Since the \oiii\ line is used extensively in this paper, we use only objects which have sufficient good pixels in its vicinity to ensure a reliable fit, as detailed in appendix A4.1 of Paper I. The 149 objects that fail this criterion are removed from the sample. The final T1 sample size is 3\,410 objects. 

The profile of the \oiii\ $\lambda5007$ emission line can be blended with the adjacent \feii\ complex and with the red wing of the broad \Hb\ profile. Therefore, we fit the wavelength region 4967\AA --5250\AA\ with three components: an iron template derived from observations of I~Zw~1 (kindly provided by T. Boroson), a GH for the \oiii\ line, and a broken power law for the underlying continuum and \Hb\ red wing. 
The measured $\loiii$ of the T1 objects are listed in Table 1. In 99\% of the objects, the relative error on the \oiii\ flux $\foiii$ is $<30\%$. In 0.5\% of the objects \oiii\ is not detected, and the quoted values are upper limits. The derivation of the error and upper limits is given in appendix B1.  

We compare our results with the $\foiii$ measurements of SDSS quasars in Shen et al. (2011), which model \oiii\ and \Hb\ using multiple Gaussians. The blending of the \oiii\ profile with the \feii\ multiplets and broad \Hb\ is pronounced mainly in luminous AGN (see fig. 18 in Paper I). Therefore, the discrepancy between different deblending algorithms in T1 objects that appear in the SDSS quasar catalog should be an upper limit on the discrepancy in the whole T1 sample. In the 419 overlapping objects, the mean ratio of our $\foiii$ to the Shen et al. $\foiii$ is 0.04 dex, with a dispersion of 0.06 dex. 

\begin{table}
{\scriptsize \begin{tabular}{l|c|c|c|c|c|c|c}
SDSS Name &  $\lbha$ & $\dv$ & $\loiii$ & $\lnha$ &  $\luv$ & $\lx$ & Notes \\ 
\hline
J000202.95-103037.9  &  41.9  &  2310  &  41.4  &  41.5  &  43.8  &  42.6  &  -,-,-,-  \\
J000410.80-104527.2  &  42.6  &  1360  &  41.5  &  41.9  &  44.6  &  43.2  &  -,-,-,-  \\
J000611.55+145357.2  &  42.1  &  3320  &  40.6  &  41.0  &  44.0  &  42.6  &  -,-,-,-  \\
J000614.36-010847.2  &  41.6  &  3910  &  40.9  &  41.1  &  43.2  &  41.9  &  -,-,-,U  \\
J000657.76+152550.0  &  41.5  &  3020  &  40.8  &  40.7  &  43.0  &  41.7  &  -,-,-,U  \\
\end{tabular}}
\caption{The broad and narrow line measurements and photometric $\nln$ luminosities of the T1 sample. All luminosities are in $\log\ \ergs$, while $\dv$ is in $\kms$. The $\luv$ and $\lx$ are measured at rest wavelengths of 1528\AA\ and 2 \kev. Notes for $\loiii,\ \lnha,\  \luv$ and $\lx$ are separated by commas, and coded as follows: `-': no note; `U': upper limit; `E': narrow line measurement relative error is $>30\%$; `O': narrow \Ha\ is \oiii-like (\S2.4.1); `A': $\lnha$ measurement is affected by stellar absorption (\S2.4.2); `N': not observed by GALEX. The electronic version includes all \smpsz\ objects.}
\label{table: sample}
\end{table}

\subsection{The narrow \Ha\ measurements}

The narrow and broad components of \Ha\ are fit simultaneously with the adjacent \oi, \nii\ and \sii\ forbidden lines, keeping the width and higher GH coefficients ($\hc,\ \hd$) of all narrow lines equal, and using a 4\th-order GH for the broad \Ha. We perform three fit attempts, in which $\hc$ and $\hd$ of the narrow lines are either zero (a pure Gaussian), same as in the $\oiii$ profile, or can vary between these two values. We use the result with the lowest $\chi^2$, and refine the fit by adding GH coefficients to the broad component. Further details of the algorithm are given in appendix A4.2 of Paper I. In appendix B2 here, we show that the three different assumptions on $\hc$ and $\hd$, and the different number of GH coefficients used for the broad component, have a negligible effect on the measured narrow \Ha\ flux, $\fnha$.

The measured $\lnha$ of the T1 objects are listed in Table 1. Two sources of uncertainty in $\lnha$ are described in the two following subsections. Additionally, objects in which only an upper limit on $\fnha$ can be measured (2\% of the sample), and objects with $>30\%$ relative error (3\%), are described in appendix B2.

\begin{figure*}
\includegraphics{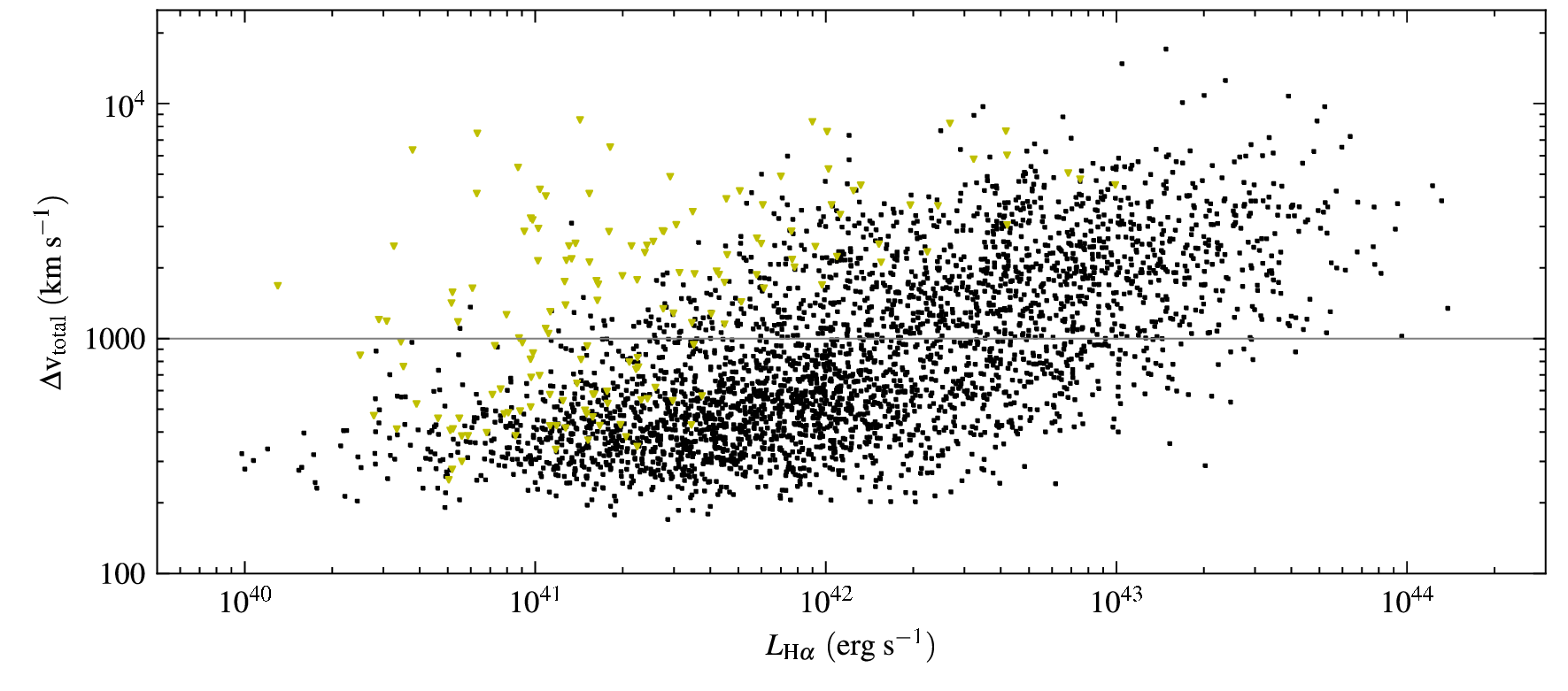}
\caption{Luminosity vs. FWHM distribution of the total (broad + narrow) \Ha\ emission line in the 3\,410 objects of the T1 sample. Yellow triangles indicate T1 objects in which $\dvt$ is overestimated due to stellar \Ha\ absorption. With decreasing $\lha$, $\dvt$ drops sharply, reaching typical NLR widths of $\sim 300\ \kms$. Almost all $\log\ \lha>43$ objects have $\dvt>1\,000\ \kms$, while most objects with $\log\ \lha<41.5$ have $\dvt<1\,000\ \kms$. Furthermore, most low luminosity high $\dvt$ objects are upper limits. The dominance of low $\dvt$ profiles at low luminosity is due to an increase in the $\lnha/\lbha$ ratio with decreasing luminosity. Therefore, selecting broad line AGN using a $\dvt > 1\,000\ \kms$ criterion (e.g. Vanden Berk et al. 2006, Schneider et al. 2010) will be highly incomplete at low AGN luminosities. 
}
\label{fig: total FWHM}
\end{figure*}

\subsubsection{\oiii -like narrow \Ha}
In 15\% of the sample, the fit yielded FWHM(n\Ha) $\geq 1.5 \times$ FWHM(\oiii). These objects have non- or barely-detectable narrow lines near \Ha, and there is no clear transition between the broad and narrow components of \Ha. In such cases, we fit the narrow \Ha\ using the FWHM, $\hc$ and $\hd$ of \oiii. 
As $\fnha$ is less certain in these objects, they are marked by different symbols in figures where $\fnha$ is used. 

It is possible that with higher spectral resolution data the demarcation between the BLR and NLR in these objects would be more clear, and one could test if the proposed fits remain consistent.

\subsubsection{Stellar absorption}

The \Ha\ absorption feature of young stars, which has EW$<5\AA$ (Groves et al. 2012), is not modeled by our fitting algorithm. Therefore, when the ratio of AGN to host luminosity is low, the measured $\fnha$ and \Ha\ flux density can be underestimated due to the underlying stellar absorption. We note that the typical absorption FWHM is $1\,300\ \kms$ (Yip et al. 2004), compared to the typical FWHM(n\Ha) of $300\ \kms$. Therefore, the error in $\lnha$ is likely $<3\AA\ \times L_{\lambda}$(host). 

We derive the galaxy continuum luminosity density in the T1 objects by subtracting the AGN contribution, using EW(b\Ha) $=570\AA$ (Paper I). In 156 (5\%) of the T1 objects, $\lnha < 3\AA \times L_\lambda$(host), implying a potentially strong error on $\lnha$ due to stellar absorption. The $\lnha$ in these objects  are marked as lower limits. Also, in these objects we mark the FWHM of the total \Ha\ profile ($\dvt$) as an upper limit, since the measured flux density of the total \Ha\ is a lower limit.

\section{The relative NLR luminosity}

Our goal is to understand the dependence of the fraction of intermediate type AGN on AGN characteristics. In Paper I, we have found that $\lbol = 130^{\times 2.4}_{\div 2.4}\ \lbha$. Thus, we use $\lbha$ as a proxy for $\lbol$, and $\lbha$ and $\dv$ to derive $\mbh$ (see eq. 2 there). 
We begin by analyzing the total \Ha\ line, i.e. the sum of the broad and narrow components, as it is less sensitive to errors in the deblending procedure. We then proceed to examining $\lnha$ and $\loiii$ directly.

\subsection{The total \Ha}

Figure 1 shows the distribution of the T1 objects in the total \Ha\ FWHM ($\dvt$) vs. total \Ha\ luminosity ($\lha$) plane. At $\lha > 3 \times 10^{43}\ \ergs$, practically all T1 objects have $\dvt>1000\ \kms$, indicating the broad component dominates the \Ha\ profile. With decreasing luminosity $\dvt$ decreases, such that 85\% of the T1s with $\lha<10^{42}\ \ergs$ have $\dvt<1\,000\ \kms$. At $\lha<10^{41}\ \ergs$, 73\% of the T1 objects have $\dvt<500\ \kms$. Widths below $500\ \kms$ are typical of the NLR, indicating that at low luminosities the NLR dominates the \Ha\ profile, or equivalently, most AGN are intermediate types. Some of the objects with low $\lha$ do have high $\dvt$, but almost all of them have a stellar absorption feature with strength comparable to the narrow emission line (\S2.4.2), indicating their $\dvt$ are upper limits, and reinforcing the observed trend of $\dvt$ with $\lha$.

In previous works, broad line AGN were selected based on optical spectra using a $\dvt > 1\,000\ \kms$ criterion (Vanden Berk et al. 2006, Schneider et al. 2010). Figure 1 implies that this criterion is adequate for quasars, but is highly incomplete at low AGN luminosities. 

\begin{landscape}
\begin{figure}
\begin{center}
\includegraphics{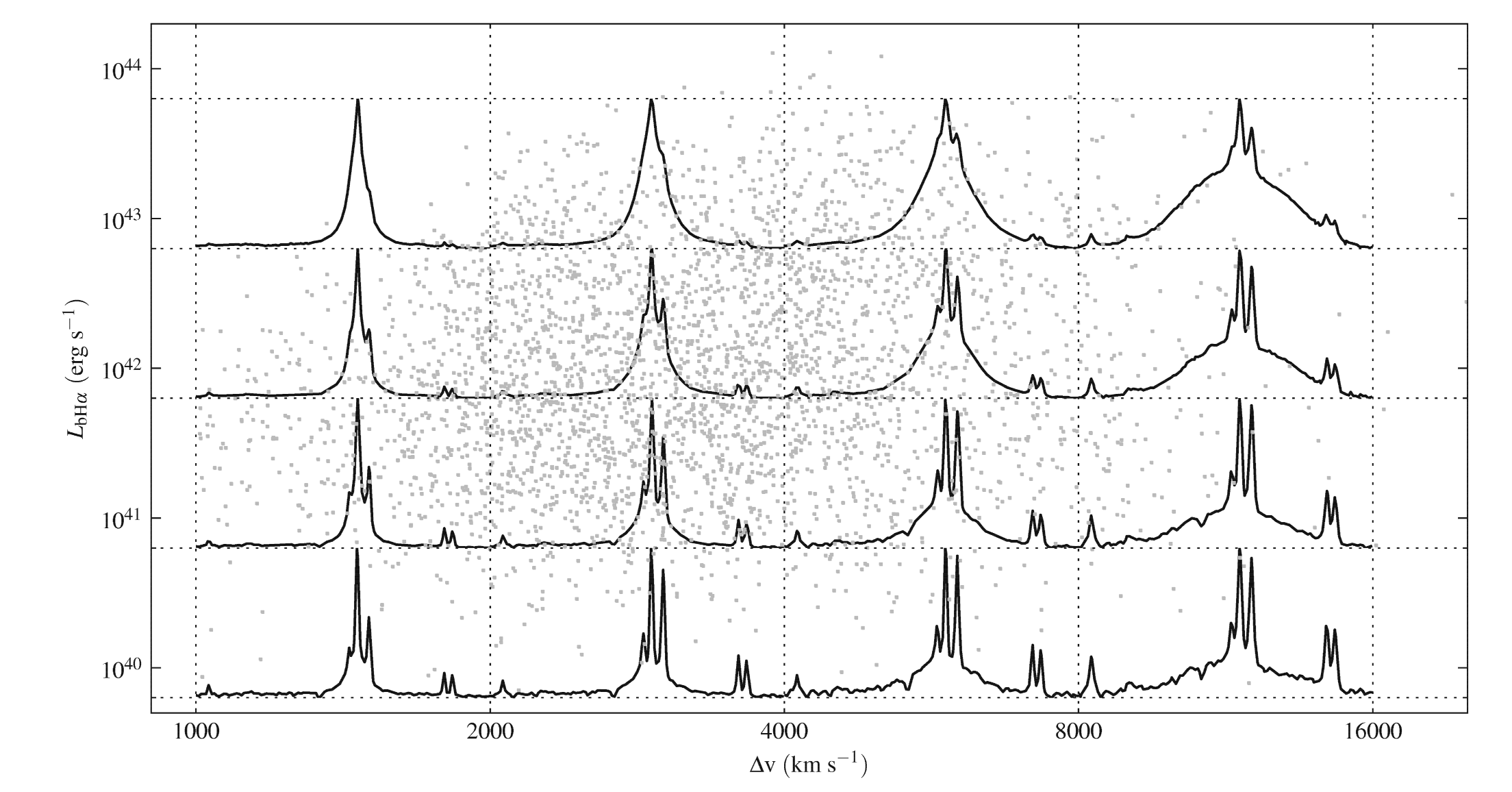}
\caption{Mean 6280\AA--6800\AA\ spectra in bins of $\lbha$ and $\dv$, overplotted on the $\lbha$ vs. $\dv$ distribution of the T1 sample. 
In each bounding box (dotted lines) we plot the mean observed spectrum, before host subtraction, of all objects with $\lbha$ and $\dv$ in the range delimited by the box. The y-axis of each spectrum is in $\ergs \AA^{-1}$, and extends from the continuum level to the max flux density of \Ha, so $\dvt$ (Fig. 1) is measured at the middle of each bounding box. 
The prominent features seen in the spectra are the broad and narrow \Ha, and the $\oi\ \lambda\lambda6300,\ 6363,\ \nii\ \lambda\lambda6548,\ 6583$ and $\sii\ \lambda\lambda6716,\ 6731$ low ionization forbidden lines. 
Note the increasing prominence of the narrow \Ha\ with decreasing luminosity, as indicated by Fig. 1. 
At $\lbha \gtrsim 10^{43}\ \ergs$ the mean \Ha\ profiles are typical of Seyfert 1.0s, at $\lbha \sim 10^{42}\ \ergs$ the mean \Ha\ profiles are typical of Seyfert 1.5s, while at $\lbha \sim 10^{40}-10^{41}\ \ergs$ type 1 AGN tend to be Seyfert 1.8s. 
The binning by $\lbha$ and $\dv$ is equivalent to binning by $\mbh$ and $\lledd$. The $\mbh$ increases upward and rightward, while $\lledd$ increases upward and leftward. At high $\lbha$ and low $\dv$, or at high $\lledd$, there is no clear transition between the narrow and broad components of the \Ha\ profiles. In such objects, we use the profile of $\oiii\ \lambda5007$ as a template for the narrow \Ha\ (\S2.4.1). 
With decreasing $\lledd$, the low ionization forbidden lines become stronger relative to the narrow \Ha, probably indicating a transition to LINER narrow line ratios (see \S3.4.3).}
\label{fig: mean spectra}
\end{center}
\end{figure}
\end{landscape}

In Figure 2, we plot the mean spectra of the T1 objects in bins of $\lbha$ and $\dv$, in the range 6280\AA--6800\AA. Each bin spans a factor of ten in $\lbha$, and a factor of two in $\dv$. The peripheral bins are not evenly populated. We therefore show in the background the distribution of the objects in the $\lbha$ vs. $\dv$ plane. Each mean spectrum is derived from all objects inside its bounding box. Mean spectra are calculated by geometrically averaging luminosity densities of spectrum pixels with the same restframe wavelength $\lambda$, rounded to $10^{-4}$ in $\log \lambda$. The spectra that enter this averaging are observed spectra, before host subtraction. The y-axis of each spectrum extends from the continuum level to the maximum flux density of \Ha, so $\dvt$ (Fig. 1) corresponds to the width of \Ha\ at the middle height of each bounding box.

With decreasing luminosity, the narrow \Ha\ component dominates the total \Ha\ profile, as indicated by the $\dvt$ analysis in Fig. 1. 
At $\lbha \gtrsim 10^{43}\ \ergs$ (top row), the mean \Ha\ profiles are dominated by the broad component, typical of Seyfert 1s. 
At $10^{43}\gtrsim \lbha \gtrsim 10^{42}\ \ergs$ (second row), the mean \Ha\ profiles indicate that most objects are Seyfert 1.5s, while at $\lbha \approx 10^{40}-10^{41}\ \ergs$ (lowest row), the objects tend to be Seyfert 1.8s. Intermediate type AGN dominate the population of broad line AGN at $\lbha<10^{42}\ \ergs$. 

Note that a relatively stronger \Ha\ stellar absorption feature at low luminosity, which is probably associated with the stronger observed stellar features, will only diminish the observed narrow \Ha\ (see appendix A). Therefore, the true trend in narrow to broad ratio is likely stronger than observed in Fig. 2.

In Fig. 2, $\mbh$ increases upward and rightward, while $\lledd$ increases upward and leftward. The dependence of the NLR to BLR luminosity ratio on $\lledd$ and $\mbh$ is examined below (\S3.4). 
Note that at high $\lbha$ and low $\dv$, or high $\lledd$, the profiles show no clear transition between the narrow and broad components of \Ha. In these objects, we fit the narrow \Ha\ with a profile identical to $\oiii$ (\S2.4.1). 
Also, with decreasing $\lbha$ and increasing $\dv$, or with decreasing $\lledd$, the low ionization forbidden lines become stronger relative to the narrow \Ha. The relative ratios of narrow emission lines, and their dependence on $\lledd$, $\mbh$, $\lbol$ and $\dv$, are the focus of a following paper.

\subsection{The narrow \Ha}

We now utilize the $\lnha$ measurements described in \S2.4. 
In Figure 3, the luminosity ratio of the narrow and broad \Ha\ is plotted versus $\lbha$. Black dots mark objects with robust $\lnha$ measurements, while other colors and shapes indicate a possible bias in $\lnha$. A trend of increasing $\lnha/\lbha$ with decreasing luminosity is clearly seen, extending over a range of $10^4$ in $\lbha$. The least-squares best fit power law of all T1 objects is
\begin{equation}
L_{\rm n\Ha;\ 42} = 0.13\ L_{\rm b\Ha;\ 42}^{0.66 \pm 0.01},\ \sigma = 0.40\ {\rm dex}
\end{equation}
where both luminosities are given in units of $10^{42}\ \ergs$, and $\sigma$ is the scatter of the measured $\log\ \lnha$ around the relation. We treat $\lbha$, which was used to select the T1 sample, as the independent variable. 

The \oiii-like narrow \Ha\ are located at the high-$\lbha$ end of the sample, with a $\lnha/\lbha$ ratio consistent with objects with robust $\lnha$ measurements. Objects with strong stellar absorption are biased to lower $\lnha$ than other T1 objects with the same $\lbha$, as expected if part of the narrow emission vanished in the \Ha\ absorption feature. Overall, the trend of $\lnha/\lbha$ with $\lbha$ is independent of the details of the \Ha\ fit, as implied by Figs. 1 and 2.

Excluding objects with \oiii-like narrow \Ha\ or strong stellar absorption, we get
\begin{equation}
L_{\rm n\Ha;\ 42} = 0.15\ L_{\rm b\Ha;\ 42}^{0.67 \pm 0.01},\ \sigma = 0.37\ {\rm dex}
\end{equation}
similar to equation 1. The measurement error on $\lnha$ is $<0.1$ dex (\S 2.4), therefore the noted dispersion is dominated by the intrinsic dispersion. 

Next, we evaluate the abundance of intermediate type AGN at different AGN luminosities. We define Seyfert 1.5s and 1.8s analytically, using the $\lnha/\lbha$ ratio. The difference between Seyfert 1.8s and 1.9s is disregarded, as it is based on the detectability of the broad \Hb. Objects with $\lnha \approx 0.1\ \lbha$, i.e. a roughly equal broad and narrow flux density for the typical width ratio of 10 ($=3\,000/300$), are considered as Seyfert 1.5s. Objects with $\lnha \approx 0.3\ \lbha$ (i.e. flux density ratio $\sim3$) are considered as Seyfert 1.8s. 

Fig. 3 shows that at $\lbha>10^{43}\ \ergs$, only 20\% of the objects are above the Seyfert 1.5 line, compared to 77\% at $\lbha<10^{42}\ \ergs$. At $\lbha<10^{41}\ \ergs$, $52\%$ of the objects are above the Seyfert 1.8 line. 
Intermediate types dominate the broad line AGN population at low luminosity, as already implied by Figs. 1 and 2.

\subsection{The \oiii\ line}

We use the \oiii\ luminosity as another measure of the NLR luminosity. Measuring the $\oiii$ line is complementary to the narrow \Ha\ measurement, as \oiii\ is less susceptible than \Ha\ to contamination from \hii\ regions powered by star formation in the host galaxy, but is more sensitive to ionization and density effects in the NLR. In Figure 4, we plot $\loiii/\lbha$ vs. $\lbha$ in the T1 sample. The least-squares best fit slope, using $\lbha$ as the independent variable, is
\begin{equation}
L_{\oiii;\ 42} = 0.16\ L_{\rm b\Ha;\ 42}^{0.72 \pm 0.01},\ \sigma = 0.36\ {\rm dex}                                            
\end{equation}
similar in slope to the $\lnha$ vs. $\lbha$ relation. The measurement error on $\loiii$ is $<0.1$ dex (\S 2.3), therefore the noted dispersion is dominated by the intrinsic dispersion. 

We add 20\,234 $z<0.82$ AGN from the SDSS quasar catalog (QCV, Schneider et al. 2010), in order to extend our luminosity range. We use the $L_{\rm b\Hb}$ measurements of Shen et al. (2011), converted to $\lbha$ using $\lbha/L_{\rm b\Hb}=3.2$, the mean ratio for 419 objects common to QCV and the T1 sample (the dispersion in $\lbha/L_{\rm b\Hb}$ is 0.1 dex). We also subtract 0.04 dex from the Shen et al. $\loiii$ (\S2.3). To prevent confusion, Fig. 4 shows only the density contours of the QCV objects. These follow the same trend as the T1 objects, up to $\lbha=5\times10^{44}\ \ergs$. The trend of increasing $\loiii/\lbha$ with decreasing $\lbha$ extends over 5 orders of magnitude in luminosity.

\begin{figure*}
\includegraphics{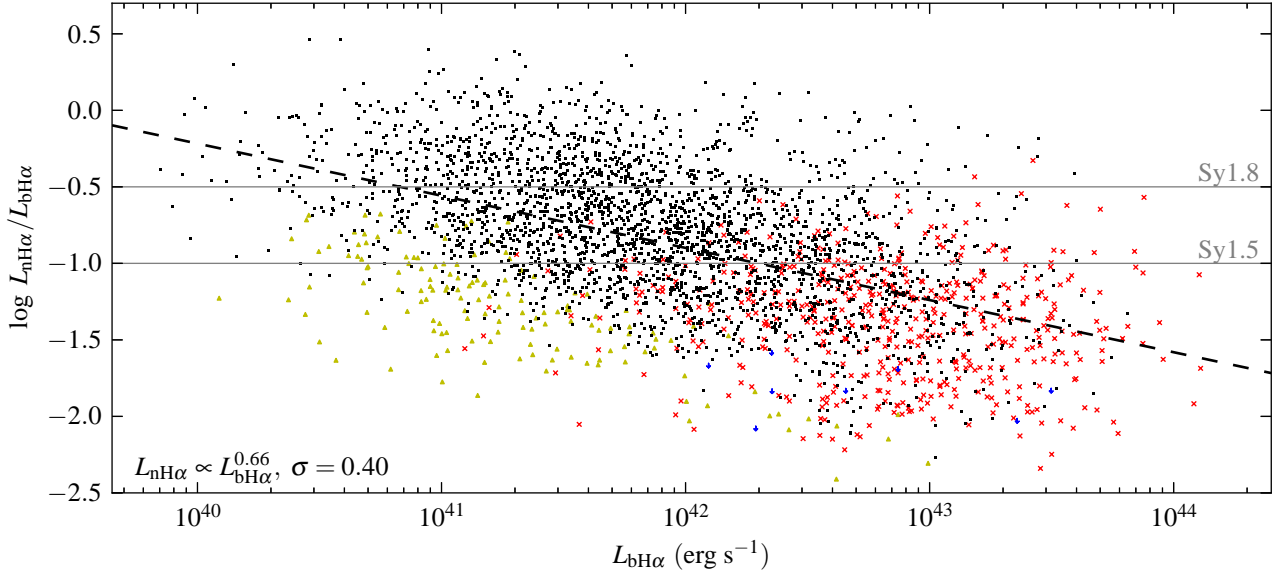}
\caption{The distribution of the NLR to BLR luminosity ratio in the T1 sample, vs. $\lbha$. Black dots mark T1 objects with robust $\lnha$ measurements, while other markers indicate the measured $\lnha$ could be biased. The first relevant uncertainty of the following list determines the shape and color of the marker: no clear BLR/NLR transition (red `x', see \S2.4.1); $\lnha$ is underestimated due to stellar absorption (yellow triangles, \S2.4.2); $\lnha$ is an upper limit (blue arrows). The slope of the best fitting power law (dashed line) and the associated dispersion are noted. The tendency of increasing NLR/BLR ratio with decreasing luminosity implied by Figs. 1 and 2 is clearly seen, extending over a range of $10^4$ in $\lbha$. Objects without a robust $\lnha$ measurement are consistent with the trend implied by objects with an accurate measurement. Objects near or above the Seyfert 1.5 line are intermediate type AGN. The abundance of intermediate types at low luminosities disfavors partial obscuration or variability as 
their origin. 
}
\label{fig: NLR BLR ratio}
\end{figure*}

\begin{figure*}
\includegraphics{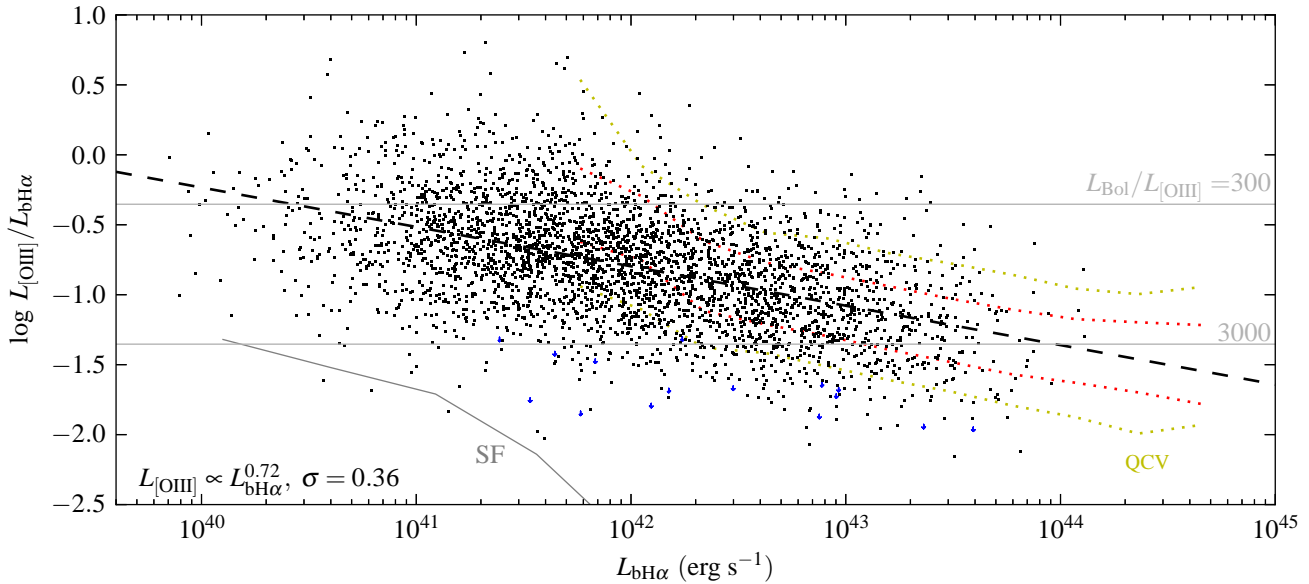}
\caption{The distribution of the \oiii\ $\lambda5007$ to broad \Ha\ luminosity ratio vs. $\lbha$, in the T1 and QCV samples. T1 objects are shown as black points, with upper limits on $\loiii$ marked by blue arrows. The slope of the best fitting power law (dashed line) and the associated dispersion are noted. The QCV distribution (measurements from Shen et al. 2011) is shown as density contours, where at each 0.3 dec wide $\lbha$ bin, 50\% of the objects lie between the inner red dotted lines, while 80\% lie between the outer yellow lines. The decrease in relative \oiii\ luminosity spans $10^5$ in $\lbha$. The implied $\oiii$ correction factor, using $\lbol = 130\ \lbha$ (Paper I) is noted. It changes by a factor of $\sim10$ over the range of luminosities shown. The mean contribution to $\loiii$ from star formation in the host galaxy (see \S5.1) is marked by `SF'. It cannot account for the observed trend, as it contributes $<10\%$ of $\loiii$. 
}
\end{figure*}

\begin{figure*}
\includegraphics{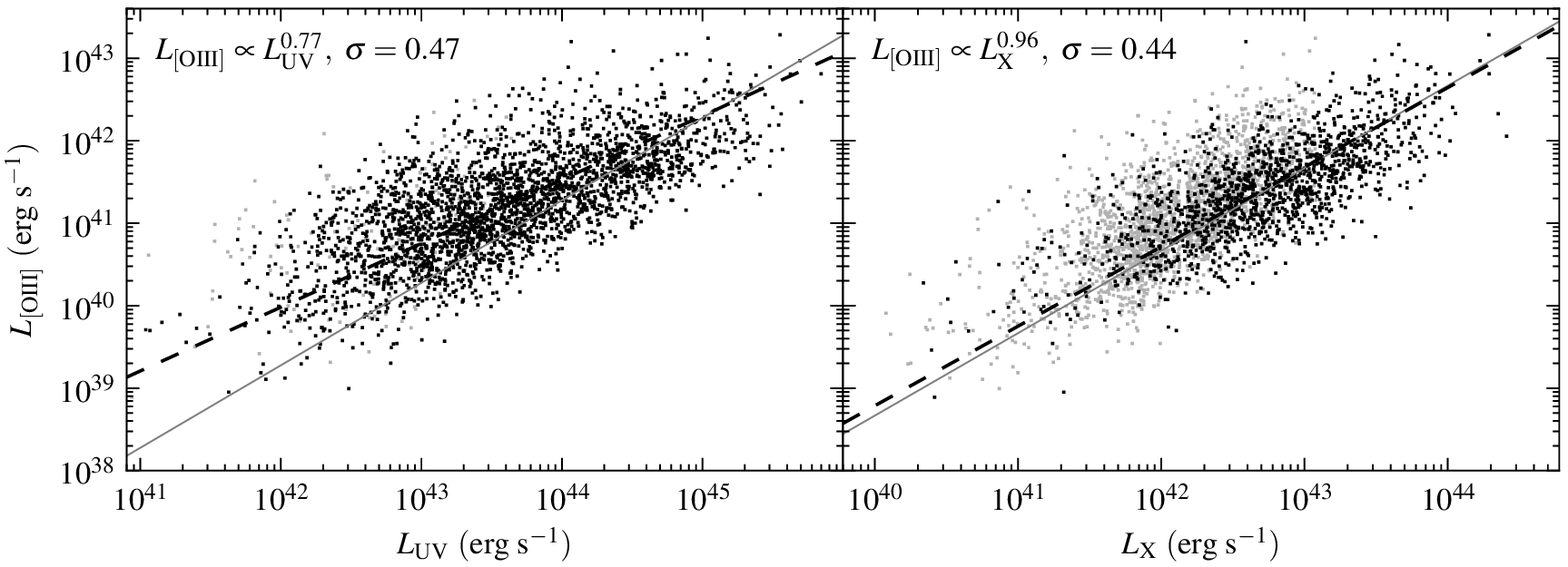}
\caption{The relation between $\loiii$ and AGN luminosity in the UV and X-ray bands, in the T1 sample. Upper limits on $\lx$ and $\luv$ are shown as gray dots. The slopes of the best fitting power laws (dashed lines) and the associated dispersions are noted. Gray solid lines depict a linear relation. {\bf (Left Panel)} The UV band is close to the peak emission of AGN, and is therefore a good indicator of $\lbol$. The linear relation is normalized by the $\loiii/\luv$ ratio in the highest $\lbha$ T1 objects. The $\loiii$ vs. $\luv$ relation is non-linear, implying an increase in $\lbol/\loiii$ with luminosity, as shown in Fig. 4. The intrinsic relation may be even more non-linear, since host emission increases $\luv$ at $\luv < 10^{43}\ \ergs$. {\bf (Right panel)} The linear relation is normalized by the $\loiii/\lx$ ratio given in Heckman et al. (2005), and is consistent with the best-fit slope. 
}
\end{figure*}

As discussed in the introduction, $\loiii$ is often used as a measure of AGN bolometric luminosity, when other measures are unavailable, particularly in obscured (type 2) AGN. Heckman et al. (2004) found $\lbol/\loiii = 3\,500$, derived from the mean EW(\oiii) of $z<0.3$ type 1 AGN. Here, we derive $\lbol/\loiii$ on the T1 sample, using the measured $\loiii/\lbha$, and the luminosity independent $\lbol/\lbha=130$ (Paper I). 
The derived relation is
\begin{equation}
\frac{\lbol}{10^{43}\ \ergs} = 4000^{\times 4}_{\div 4}\ (\frac{\loiii}{10^{43}\ \ergs})^{1.39}
\end{equation}
Derivation of the uncertainty is described below. 

The implied $\lbol/\loiii$ values are indicated by two horizontal lines in Fig. 4. The mean $\lbol/\loiii$ drops from 3\,000 at $\lbha=10^{44}\ \ergs\ ( \lbol =10^{46.1}\ \ergs)$, to 300 at $\lbha = 2\times10^{40}\ \ergs\ ( \lbol =10^{42.5}\ \ergs)$. At the low luminosity end of the T1 sample, our mean $\lbol/\loiii$ is smaller by a factor of $>10$ from the Heckman et al. result. We elaborate on this discrepancy in \S4. 

In Figure 5, we compare $\loiii$ with $\luv$ and $\lx$, which are other measures of the bolometric luminosity. The best fit for the 93\% of the T1 objects with detections in the UV (\S2.1.1) is
\begin{equation}
L_{\oiii;\ 42} = 0.01\ L_{\rm UV;\ 42}^{0.77 \pm 0.01},\ \sigma = 0.47\ {\rm dex}
\end{equation}
As neither $\loiii$ nor $\luv$ were used in the selection of the T1 sample, we derive the best fit by minimizing the 2D distances of the data points from the fit. The noted $\sigma$ is the scatter in the distance of $\log\ \loiii$ from the best fit relation, as in eqs. 1--3. 

The $\loiii$ vs. $\luv$ slope is similar to the product of the linear $\lbha$ vs. $\luv$ relation (Paper I), and the 0.72 slope of $\lbha$ vs. $\loiii$ (eq. 3). Note that the host contribution to the UV becomes non-negligible at $\log\ \lbha < 42$, and can bias eq. 5. A similar best fit, using only T1s with $\log\ \lbha > 42$, gives a slope of 0.65. 
Since the UV emission occurs close to the position of the SED peak in AGN (e.g. Zheng et al. 1997), eq. 5 emphasizes the non-linear relation between $\loiii$ and $\lbol$ found above. 

In eq. 4 we use the 0.6 dex scatter of $\log\ \luv$ around the $\luv$ vs. $\loiii$ relation, as an estimate for the uncertainty in deriving $\lbol$ from $\loiii$. Measurement errors are $\sim 0.02$ dex in $\luv$ (Morrissey et al. 2007), and $<0.1$ dex in $\loiii$, as noted above. Therefore, the uncertainty is dominated by the physical dispersion. Moreover, extinction along the line of sight probably has a significant effect on $\luv$ (\S3.7.3 in Paper I), indicating the true dispersion between $\loiii$ and $\lbol$ might be lower.

A similar comparison of $\loiii$ and $\lx$, on the 43\% of the T1 objects with ROSAT detections, gives
\begin{equation}
L_{\oiii;\ 42} = 0.05\ L_{\rm X;\ 42}^{0.96 \pm 0.02},\ \sigma = 0.44\ {\rm dex}
\end{equation}
The coefficient and dispersion are consistent with the $\loiii/L_{\rm 2-10\ \kev}$ ratio of 0.03 and dispersion of 0.48 dex given in Heckman et al. (2005), found for type 1 AGN selected by their \oiii\ emission (we converted $L_{\rm 2-10\ \kev}$ to $\lx$ using $\alpha_{\rm x}=-1.5$, \S2.1.2). The linear relation is not surprising, given the similar slopes of $\loiii$ and $\lx$ vs. $\lbha$ (eq. 3 here and eq. 7 in Paper I). Also, the known $\lx \propto L_{\rm 2500\AA}^{\sim0.7}$ relation (e.g. Just et al. 2007), is similar in slope to the $\loiii$ vs. $\luv$ relation (eq. 5) found here,  which also points to a relation between $\loiii$ and $\lx$ which is close to linear. Eq. 6 could be biased due to the low X-ray detection fraction of 43\% (\S2.1.2). For objects with $\fbha >10^{-13.5}\ \flux$, where the X-ray detection fraction is $61\%$, the slope is $0.91\pm0.03$. Note that previous studies of type 1 AGN found flatter $\loiii$ vs. $\lx$ slopes: $0.7 \pm 0.06$ in Netzer et al. (2006), $0.82 \pm 0.04$ in Panessa et al. (2006) and $0.68 \pm 0.2$ in Trouille \& Barger (2010).

\subsection{The correlation of $L_{\rm NLR}/L_{\rm BLR}$ with $\lledd$ and $\mbh$}

In the previous sections, we find a trend of decreasing $\lnha/\lbha$ and $\loiii/\lbha$ with increasing $\lbha$. Here we explore which of the parameters, $\lbha$, $\mbh$ or $\lledd$, appears to be the dominant parameter which drives the $L_{\rm NLR}/L_{\rm BLR}$ trends.

\subsubsection{The T1 sample}

We bin the T1 objects based on $l\ (\equiv \log\ \lledd)$ and $m\ (\equiv \log\ \mbh/\msun)$ in the following manner. The objects are sorted by $m$ and divided into six equal size groups. Each of these groups is then sorted by $l$, and again divided into six equal size groups. This ensures similar statistical errors in all bins. We repeat this process by sorting based on $\lbha$ and $\dv$. 

Figure 6 presents the derived relations of the mean values of $\lnha/\lbha$ and $\loiii/\lbha$ as a function of $\lbha$, at a fixed $\dv$ (left columns) or as a function of $m$, at a fixed $l$ (right columns). Error bars denote the uncertainty in the mean. In the upper row, objects with strong stellar absorption are disregarded since they are offset to lower $\lnha/\lbha$ values (Fig. 3). The upper-left panel shows that $\lnha/\lbha$ decreases with $\lbha$, as shown in Fig. 3. The dependence of $\lnha/\lbha$ on $\dv$ is weak, if at all. The upper right panel shows that both $m$ and $l$ contribute to the trend, as $\lnha/\lbha$ decreases with increasing $m$, at a fixed $l$, and also with increasing $l$, at a fixed $m$. The comparison of the upper two panels indicates that $\lbha$ (or $\lbol$) is the main driver of the $\lnha/\lbha$ ratio, and the trends with $m$ and $l$ are driven by the $\lbha$ dependence. 

\begin{landscape}
\begin{figure}
\begin{center}
\includegraphics{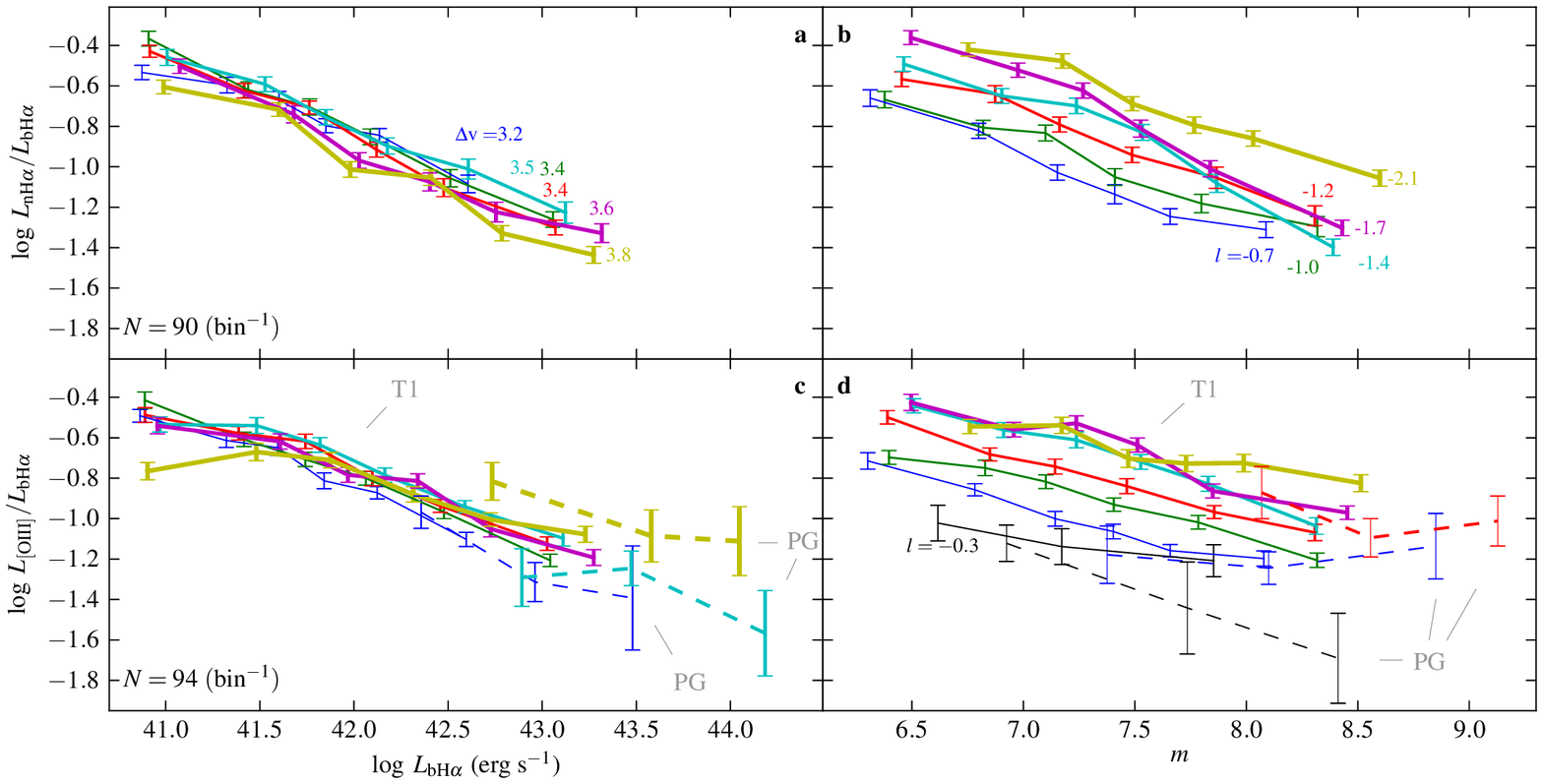}
\caption{Mean NLR to BLR emission ratios in the T1 sample, as a function of $\lbha$ and $\dv$ (left column), and as a function of $m\ \equiv \log \mbh/\msun$ and $l\ \equiv \log \lledd$ (right column). All T1 bins in each row have the same number of objects, as noted in the lower left corner. Same $\dv$ or $l$ bins are connected by solid lines, with thickness increasing with $\dv$ or decreasing $l$ (mean $\dv$ or $l$ noted). Error bars denote the uncertainty in the mean value. 
{\bf (a, b)} Objects with strong stellar absorption (\S2.4.2) are disregarded. The mean $\lnha/\lbha$ is largely set by $\lbha$, and independent of $\dv$. The $\lnha/\lbha$ decreases both with $m$ at a fixed $l$ and with $l$ at a fixed $m$, indicating that $\lbha$ (or $\lbol$) is the key parameter driving the $\lnha/\lbha$ trend. 
{\bf (c, d)} 
The mean $\loiii/\lbha$ is set by $\lbha$, and independent of $\dv$, similar to $\lnha/\lbha$. Also plotted is the PG sample, divided into bins with nine objects each (values from Baskin \& Laor 2005a, 2005b). The PG bins are connected by dashed lines, with color and thickness matched to the T1 bins. The black line shows the 42 T1 objects with $l \sim -0.3$, divided into three $m$ bins. The PG sample shows an increase of $\loiii/\lbha$ with $\dv$ at a fixed $\lbha$, as implied by EV1 in Boroson \& Green (1992), and in contrast with the T1 sample. On the other hand, the PG $\loiii/\lbha$ is consistent with T1 at $m<8.5$, while the trend of decreasing $\loiii/\lbha$ with increasing $m$ reverses at $m>8.5$. There are few $m>8.5$ T1s, thus inducing the apparent discrepancy in the left panel. The reversed trend of $\loiii/\lbha$ is due to the dominance of radio loud AGN at $m>8.5$. The $7\sigma$ difference in $\loiii/\lbha$ between the $\log \dv=3.8$ bin and other $\dv$ bins at the lowest $\lbha$, corresponds to the Seyfert / LINER transition at $l \sim -2.7$ found by Kewley et al. (2006).
} 
\label{fig: o3 bol}
\end{center}
\end{figure}
\end{landscape}

In the lower panels, the analysis is repeated for $\loiii/\lbha$. The behavior of $\loiii/\lbha$ with AGN characteristics is similar to the behavior of $\lnha/\lbha$. The trend of decreasing $\loiii/\lbha$ with increasing $m$ or $l$ are driven by the $\lbha$ dependence. The $\loiii/\lbha$ is dependent on $\dv$ only at the lowest $\lbha$ (see below). 

We conclude that the decrease in relative NLR luminosity in the T1 sample is mostly due to an increase in AGN luminosity, and not due to an increase in $\lledd$ or $\mbh$. This result apparently contradicts earlier results, e.g. for the PG sample (Boroson \& Green 1992). We therefore repeat our analysis on the PG sample, as described below.

\subsubsection{The PG sample}

Boroson \& Green (1992) performed a PCA on the PG quasar sample (Neugebauer et al. 1987). Their first eigenvector (EV1) indicates that $\loiii/\lbhb$ correlates with FWHM(\Hb), while their EV2 indicates that $\loiii/\lbhb$ anti-correlates with the absolute V-magnitude. In Boroson (2002), EV1 was interpreted as an indicator of $\lledd$, while EV2 follows $\lbol$. As seen in Fig. 6, in the T1 sample we recover the EV2 trend of $\loiii/\lbha$ with $\lbol$, but do not see a trend of $\loiii/\lbha$ with $\dv$, apparently discrepant with EV1.

We add the 81 $z<0.5$ PG quasars to Fig. 6, divided into $3\times3$ bins and color coded as the T1 sample. We use $\loiii$, $\lbhb$, and FWHM(\Hb) from Baskin \& Laor (2005a, 2005b), assuming FWHM(\Hb)$=\dv$ and $\lbha(\rm T1) / \lbhb({\rm PG}) = 2.45$ (calculated from common objects). 
The PGs have a bin with $l=-0.3$, which does not appear in T1. Therefore, we also plot the positions of the 42 T1 objects with $l \sim -0.3$, divided into three $m$ bins (these are the 42 objects with the highest $l$ in the $l=-0.7$ bin of T1).

In the lower left panel, it can be seen that $\loiii/\lbha$ increases by a factor of up to $\sim 2.5$ over a factor of 4 in $\dv$ at a given $\lbha$, as implied by the EV1 relations, and in contrast to the lack of trend in $\loiii/\lbha$ vs. $\dv$ in the T1 sample. In the lower right panel, it can be seen that $\loiii/\lbha$ in the PG sample is consistent with the T1 relations at $m<8.5$. However, the trend of decreasing $\loiii/\lbha$ with increasing $m$ reverses at $m>8.5$. 
We note that only 5\% of the T1s have $m>8.5$, compared to 33\% of the PG quasars.

What can cause this change of trend in $\loiii/\lbha$ at high $\mbh$? 
A possible suspect is radio loudness, since strong radio emission is known to be associated with enhanced narrow line emission (de Bruyn \& Wilson 1978, and citations thereafter) and with large $\mbh$ (Laor 2000). Specifically, 52\% of the $m>8.5$ PG quasars at $z<0.5$ (i.e. the Boroson \& Green sample) are radio loud, compared to 2\% of the $m<8.5$ PGs (fig. 2 in Laor 2000). Therefore, we suspect that emission line filaments associated with radio lobes in radio loud quasars provide additional sites for \oiii\ emission at high $m$, and drives the $\loiii/\lbhb$ vs. $\dv$ correlation of EV1. At $m<8.5$, the dominant trend of $\loiii/\lbha$ in the PG sample is with $\lbol$, as found for the T1 sample.

\subsubsection{LINERs}

Kewley et al. (2006, hereafter K06), showed that SDSS type 2s occupy two `clouds' in the BPT diagrams (Baldwin et al. 1981, Veilleux \& Osterbrock 1987), such that LINERs (Heckman 1980) have distinctly weaker $\loiii/\lnhb$ and stronger $L_{\sii}/\lnha$ and $L_{\oi}/\lnha$ than Seyferts. 
K06 showed that Seyferts and LINERs are separated by a threshold of $\loiii/\sigma_*^4 \sim 10^{-1.9}\ L_{\odot}\ (\kms)^{-4}$. 
This threshold is equivalent to $l = -2.7$, using the G{\"u}ltekin et al. (2009) $\mbh-\sigma_*$ relation, and eq. 4 at $\loiii=10^{6.8}\ L_\odot$, the typical $\loiii$ at the Seyfert/LINER transition (fig. 16 in K06). Do we see this transition in the T1 sample?

In the lower left panel of Fig. 6, the mean $\loiii/\lbha$ is lower by a factor of two (difference of 7$\sigma$) in the $\log\ \lbha=41,\ \log \dv=3.8$ bin, compared to other bins with the same $\lbha$. Objects in this bin have $l = -2.6$, the lowest $l$ in the T1 sample\footnote{Note that due to our binning method, the lowest $l$ bins in the upper right panel have a larger mean $l \sim -2.2$, and do not show a similar drop.}, and similar to the threshold $l$ found by K06 for the Seyfert/LINER transition. Therefore, the drop in $\loiii/\lbha$ is probably related to the LINER phenomenon. The LINER interpretation is also supported by the high $L_{\oi}/\lnha$ and $L_{\sii}/\lnha$ ratios in the mean spectra of these objects (Fig. 2). 
Thus, the $\loiii \propto \lbol^{0.7}$ relation found above is probably applicable only at $l > -2.6$, for AGN classified as Seyferts.

\section{Comparison with previous studies}

We find here that at $\lbol=10^{46}\ \ergs$, $\lbol/\loiii=3\,000$; while at $\lbol = 10^{42.5}\ \ergs$, $\lbol/\loiii=300$ (Fig. 4).  
At low AGN luminosity, our $\lbol/\loiii$ differs by a factor of $>10$ from results of previous studies, such as the $\lbol/\loiii=3\,500$ found by Heckman et al. (2004, hereafter H04). In this section, we compare our methods with prior work, to explore the reason for the discrepancy.

Croom et al. (2002), H04 and Netzer et al. (2006), found that the mean EW(\oiii) remains constant with luminosity in type 1 AGN samples with $z<0.3$, which should overlap well the T1 sample. At higher $z$ and higher AGN luminosity, Sulentic et al. (2004) and Netzer et al. (2006) found a decrease of EW(\oiii) with $\lbol$, dubbed the `disappearing NLR'. This transition between the flat and decreasing parts of the EW(\oiii) vs. $L_{5100\AA}$ relation can be seen in fig. 9 of Ludwig et al. (2009). 

In Paper I, we derived the mean $\lhost/\lagn$ ratio at 5100\AA, for different $\lbol$ (fig. 13 there). At $\log\ \lbol = 45.5$, $\lhost~/~\lagn \sim 0.2$, while at $\log\ \lbol= 43$, $\lhost~/~\lagn \sim 10$. Therefore, the host contribution to the continuum increases from $\sim 15\%$ to $\sim90\%$ over a range of $10^{2.5}$ in $\lbol$. This increase in host contribution can also be seen in the decrease of stellar absorption features EW with luminosity (fig. 8 in Croom et al 2002; fig. 18 in Paper I). Hence, if the host is not accounted for, at $\log\ \lbol=43$ the measured EW(\oiii) will be lower by a factor of $\sim10$ than its intrinsic value. Above $\log\ \lbol = 45.5$, where the host contribution at 5100\AA\ is negligible, the measured EW(\oiii) becomes equal to the intrinsic value. None of the studies mentioned above accounted for the host contribution to the continuum when measuring EW(\oiii). Therefore, at low luminosities (or low $z$), the increase in relative host contribution with decreasing $\lbol$ cancels the intrinsic increase in EW(\oiii), yielding the measured apparent 
constant EW(\oiii) vs. $\lbol$ relation found in earlier studies. At high $\lbol$ or $z$, the host contribution is negligible, and the intrinsic decrease in EW(\oiii) emerged. Here, we use $\lbha$ as a bolometric indicator, which is not subject to host contamination, and therefore the intrinsic decrease in EW(\oiii) is apparent at all luminosities.

\section{The physical mechanism behind the $L_{\rm NLR}/\lbol$ vs. $\lbol$ trend}

In \S3 we have established a trend of decreasing $L_{\rm NLR}/L_{\rm BLR}$ with increasing $L_{\rm BLR}$, or equivalently, a dominance of intermediate type AGN at low $\lbha$. In this section, we explore the possible physical mechanisms that drive this trend.

Changes in the ionization and density of the NLR gas will have a weak effect on the \Ha\ recombination line, which is emitted proportionally to the ionization rate in the low density NLR gas (e.g. Osterbrock \& Ferland 2006). Therefore, such changes cannot drive the trend in $\lnha/\lbol$. Evidence supporting or opposing other possible physical mechanisms is presented in the following subsections.

\subsection{Star formation}

Narrow line emission can come from \hii\ regions in the host galaxy, powered by ionizing radiation from young OB stars. Can a relative increase in star formation (SF) with decreasing $\lbol$ explain the increase in $L_{\rm NLR}/L_{\rm BLR}$ by a factor of 10? The SF contribution to $\oiii$ emission in high metallicity galaxies is relatively weak, and AGN hosts generally have high metallicities (Groves et al. 2006, and references therein). Specifically, Moustakas et al. (2006) show that the expected $\loiii/\lnha$ from SF is 0.07, in contrast to the T1 sample which has $\loiii/\lnha = 1.5$ at $\log\ \lbha=43.5$, and $\loiii/\lnha = 0.9$ at $\log\ \lbha=40$. Therefore, SF does not dominate the narrow line emission at low AGN luminosity. The expected contribution to $\loiii/\lbha$ from SF, based on the host UV emission in the T1 sample and the $\loiii/\luv$ ratio of SF galaxies, is shown in Fig. 4. Its derivation is given in appendix C.

\subsection{Optically thin dust}

A possible interpretation for intermediate type AGN is that the central source is viewed through optically thin dust close to the
center, which extincts the AGN continuum and BLR, but has a negligible effect on the NLR, which originates from larger scales. Dust extinction has been found in intermediate type AGN (Goodrich et al. 1995, Maiolino \& Reike 1995, Trippe et al. 2010), and also in the T1 sample (Paper I). Higher dust column densities, which obscure the broad optical Balmer lines, but reveal broad IR Paschen and Brackett lines have also been found in AGN (Veilleux et al. 1997). Therefore, a range of dust opacities exists in AGN. Can this range explain the trends seen in Figs. 3 and 4? 

We test this scenario in Figure 7, which compares $\luv/\lbha$ with $\lbha/\lnha$ in the T1 sample. If optically thin dust drives the range in $\lbha/\lnha$, we should see a relation between $\luv/\lbha$ and $\lbha/\lnha$, as most dust models have larger opacity in the UV than near \Ha\  (e.g. Laor \& Draine 1993). The expected slope of the correlation for Milky Way dust is shown in the plot\footnote{To calculate that effect of dust on $\luv$, we assume an unabsorbed UV slope of $-0.3$, apply the extinction law, redshift by the T1 sample mean $z$ of 0.13, and convolve the result with the response functions of the GALEX filters. Our conclusions do not change for a different reasonable $z$ or intrinsic UV slope.}. 
Clearly, the observed range of $\luv/\lbha$ can explain only a small range in the observed 
$\lbha/\lnha$ values. Other dust compositions, such as those found in the SMC and LMC, have steeper extinction curves, and therefore
create a smaller range in $\lbha/\lnha$ for the observed range in $\luv/\lbha$. Hence, they are even more discrepant with the data. This result coincides with the conclusion of Trippe et al. (2010) that most intermediate type AGN are not viewed through optically thin dust.

\begin{figure*}
\includegraphics{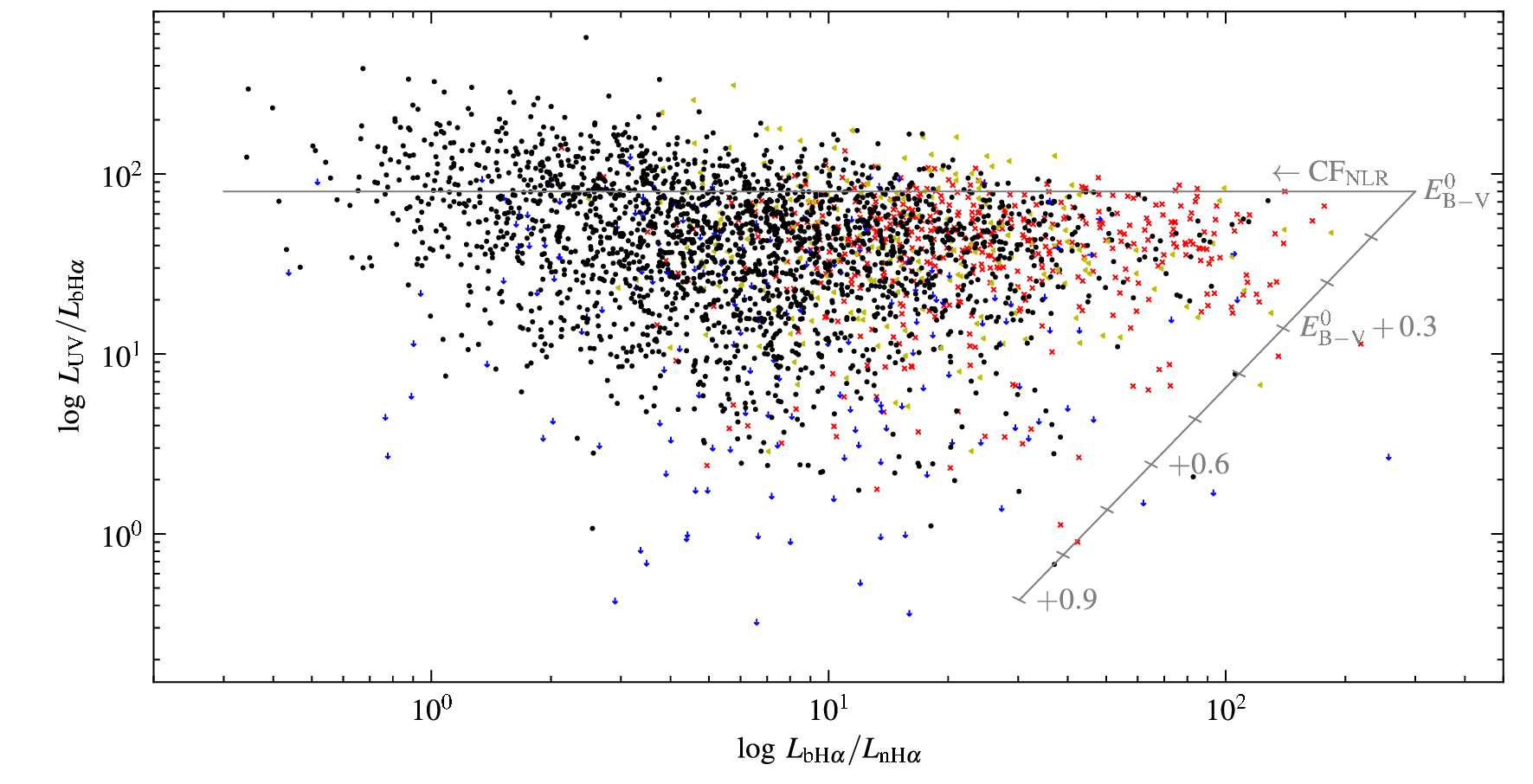}
\caption{The effect of reddening and NLR covering factor on $\luv/\lbha$ vs. $\lbha/\lnha$ in the T1 sample. Upper limits on $\luv$ are marked by down-pointing blue arrows. The symbols of other T1 objects are based on the measurement of $\lnha$, as described in Fig 3. 
The diagonal gray line depicts the effect of modifying the column density of dust along the line of sight to the central source (marks note steps of 0.1 in $\ebv$, starting from an arbitrary value $\ebv^0$). The horizontal line depicts the effect of modifying $\cfnlr$. The observed range of $\luv/\lbha$ indicates that optically thin dust can explain only a small part of the range in $\lbha/\lnha$. The range of $\lbha/\lnha$ in the T1 sample is most likely dominated by a range of $\cfnlr$. 
}
\label{fig: total FWHM}
\end{figure*}

\subsection{Partial Obscuration}

Another mechanism that can reduce the observed BLR emission is an optically thick obscurer that partially covers the BLR. The following quantitative argument can be applied to decide whether partial obscuration is a likely mechanism behind the majority of intermediate types. 

A partial obscuration of the BLR in a significant number of objects requires an absorbers size, $r_{\rm abs}$, which is
comparable to the size of the BLR, $r_{\rm BLR}$. The typical $\Delta v\sim 3000$~km~$^{-1}$, implies that the
BLR arises from a distance of $r_{\rm BLR}\sim 10^4\ G\mbh/c^2$ from the center, while the accretion disk UV continuum arises from
a distance of $r_{\rm UV} \sim 10-100\ G\mbh/c^2 $. Thus, $r_{\rm abs}/r_{\rm UV}\sim 100-1000$, and the absorber will either
completely absorb the UV source, or not absorb it at all. For example, if 90\% of the BLR is covered, for most configurations of the obscurer we expect 90\% of the objects not to be detected in the UV. This is clearly not seen in Fig. 7. Objects with $\lbha/\lnha \sim 30$ have $94\%$ UV detections, while objects with $\lbha/\lnha \sim 3$ have 92\% UV detections.  Therefore, it is unlikely that the trend in $\lnha/\lbha$ vs. $\lbha$ is due to partial obscuration.

\subsection{Variability}

The BLR samples the AGN emission on timescales $\gtrsim1\,000$ times shorter than the NLR (e.g. Laor 2003). Therefore, if an AGN is observed at a temporarily low state, it can appear as an intermediate-type AGN. However, such a scenario would imply that intermediate types should be relatively rare (to keep the mean luminosity high), contrary to the results of Figs. 3 and 4. Therefore, variability cannot explain the observed trend in $\lnha/\lbha$ vs. $\lbha$.

\subsection{Reddening in the NLR} 

Kauffman et al. (2003) find a correlation between NLR reddening and AGN luminosity in type 2 AGN (fig. 21 there). They derive the amount of reddening from the Balmer decrement, and the AGN luminosity from $\loiii$, corrected for the reddening. Could reddening of the NLR at high luminosity type 1 AGN, which does not affect the BLR, explain the drop in $\lnha/\lbha$ at high $\lbha$? 

If the extinction in the NLR comes from a dust screen external to the NLR clouds, it is unlikely that the average BLR is not affected, as the BLR originates from a region $\sim 10^3$ times smaller than the NLR, as mentioned above. Alternatively, emission line photons may be absorbed by dust embedded in the NLR clouds, which do not reside along the line of sight to type 1 AGN. However, such dust is also unlikely to explain the trends seen in Figs. 3 and 4, since $\loiii$ and $\lnha$, which have different dust opacities, show a similar trend with $\lbha$.

\subsection{Covering factor}

Another mechanism that can explain the decrease in the NLR to BLR luminosity ratio with $\lbha$ is a decrease in $\cfnlr$, or an increase in the BLR covering factor ($\cfblr$). In Paper I, we showed that the mean $\cfblr$ remains constant with $\lbha$, and therefore cannot explain the observed trend. As all other mechanisms suggested in \S\S5.1 -- 5.5 are unlikely, we conclude that the drop in $\lnha/\lbha$ and $\loiii/\lbha$ with $\lbha$ is most likely due to a decrease in $\cfnlr$ with AGN luminosity. A change in $\cfnlr$ will not affect $\luv/\lbha$, as observed in the T1 sample (Fig. 7). This result is consistent with the result derived by Baskin \& Laor (2005b) for the PG quasar sample, based on photoionization modeling of the relative strengths of [O~III]~$\lambda 5007$, [O~III]~$\lambda 4363$, and the narrow H$\beta$ line. A similar suggestion was made by Ludwig et al. (2009). Below we derive the values of the $\cfnlr$ at different $\lbol$ (\S6.4).

\section{Implications}

\subsection{Intermediate type AGN}

Above we found that at $\lbol<10^{44}\ \ergs$, the majority of type 1 AGN are intermediate types. The abundance of intermediate type AGN at low $\lbol$ is consistent with the result of Ho et al. (1997), that at $\lbha \sim 10^{39}\ \ergs$ the $\lbha$ is typically $\sim30\%$ of the total \Ha+\nii\ luminosity. 

Consequently, Seyfert 1.5 samples will tend to have lower intrinsic luminosities than Seyfert 1s, while Seyfert 1.8s will have even lower $\lbol$. The tendency of Seyfert 1.8s to have low $\lbol$ has been noted by Deo et al. (2007, 2009) and Trippe et al. (2010). At luminosities below those spanned by the T1 sample, the broad lines may be too weak to be detectable, and unobscured AGN will be misclassified as obscured AGN (see next section). 

Also, we show in \S5.2 that the inner region of low luminosity intermediate type AGN is not viewed through optically thin dust. Therefore, low luminosity intermediate type AGN should be generally treated as low luminosity unobscured AGN, and not grouped as likely obscured type 2s. The unobscured nature of most low luminosity intermediate types has been noted previously by Barth (2002).

\subsection{The $\loiii$ vs. $\lbha$ relation as an indicator of true type 2 AGN}

Various theoretical models of the BLR predict the existence of `true type 2' AGN, in which the BLR does not exist (Nicastro 2000, Elitzur and Shlosman 2006), or is not photoionized (Laor \& Davis 2011), as opposed to standard type 2 AGN, in which the BLR is obscured. Candidates are often found by looking for X-ray sources indicating nuclear activity, with no sign of obscuration in the X-ray spectrum, and no sign of a BLR in their optical continuum (Rigby et al. 2006, Trump et al. 2009, Shi et al. 2010, Tran et al. 2011).

In paper I we used the $L_{\rm X}$ vs. $\lbha$ relation to test whether the lack of detection of broad lines is indeed significant, and justifies the true type 2 identification. Here we describe how the $\loiii$ vs. $\lbha$ relation (eq. 3) can also be used for the same purpose. We compare the upper limits on $\lbha$ of six true type 2 candidates with their predicted $\lbha$ based on their observed $\loiii$. 

We use the three objects in Tran et al. (2011), NGC~4594 and IRAS~01428--0404 from Shi et al. (2010), and Q2131-427 from Panessa et al. (2009), since they all have measurements of $\loiii$\footnote{The $\loiii$ of IRAS~01428--0404 is derived from the 6dF spectrum (Jones et al. 2004, 2009).} and of the continuum flux density near \Ha, and estimates of $\mbh$ and $\lledd$. For each object we set an upper limit on the broad \Ha\ flux equal to 10\% of the continuum flux density near \Ha\ \footnote{In NGC 4450 and NGC 4579, HST detected broad \Ha\ with flux densities which are $\sim10\%$ of the stellar continuum in $2\arcsec \times 4\arcsec$ ground based measurements, where the broad \Ha\ is invisible (Ho et al. 2000, Barth et al. 2001).} multiplied by the expected $\dv$. The expected $\dv$ is estimated from 
\begin{equation}
 \dv = 1\,850\ (\frac{\mbh}{10^8\ \msun})^{0.24} (\frac{L}{L_{\rm Edd}})^{-0.24}~~ \kms
\end{equation}
which is derived from eqs. 2 and 3 in Paper I. We neglect the dependence on $\lbha$, which has an index of $1/45$.

The six objects are plotted in Fig. 8, using the measured $\loiii$ and derived upper limit on $\lbha$. The T1 sample is shown in the background, together with eq. 3 and the associated dispersion.
Except Q2131-427, all candidates have upper limits which are above their expected $\lbha$. The Q2131-427 upper limit is 2.5$\sigma$ lower than the expected $\lbha$, but still within the distribution spanned by the T1 sample. Thus, the absence of a broad \Ha\ in these six objects is not highly significant in any of the objects. A high angular resolution spectrum is required to exclude the strong host contribution near 7000\AA, and to be able to detect or exclude the expected weak \Ha. Such weak \Ha\ features have been detected from the ground using a $1\arcsec \times 1\arcsec$ aperture (Barth 2002) and with HST (Ho et al. 2000, Barth et al. 2001).

In Paper I, we performed a similar analysis using the $\lx$ vs. $\lbha$ relation (fig. 14 there). 1ES~1927+654 and IRAS~01428--0404 were found to have upper limits on $\lbha$ which are well below the expected $\lbha$ based on their $\lx$, suggesting they are potentially true type 2 AGN. These objects have high values of $\lx/\loiii=720$ (1927+654) and $230$ (01428--0404), which are $3.5$ and $2.4\sigma$ above the observed mean ratio in the T1 sample (eq. 6). These unusually high values probably result from unusually low $\loiii$, most likely due to obscuration or 
absorption. This may therefore hint that 1ES~1927+654 and IRAS~01428--0404 are obscured AGN, rather than true type 2s. Obscuration of the NLR must be produced by galactic scale dust. Such extended dust can have a column of $\sim 10^{22}$~cm$^{-2}$, sufficient to extinct the optical line emission, but be completely thin to the X-ray emission near 1 \kev. 

Thus, $\loiii$ may provides a significant additional constraint on the nature of AGN, which may appear to be true type 2, based on their $\lx$ and $\lbha$ upper limit. We note in passing that the factor of $\sim 10$ smaller column required to absorb the $5000\AA$ emission, compared to the
$\sim 1\ \kev$ emission (e.g. Laor \& Draine 1993), allows in principle to produce type 2 AGN where the optical BLR 
emission is absorbed, but the absorber is transparent to the X-ray emission. Only in the mid IR, the dust opacity becomes
low enough that the absence of X-ray absorption implies no absorption.

\begin{figure}
\includegraphics{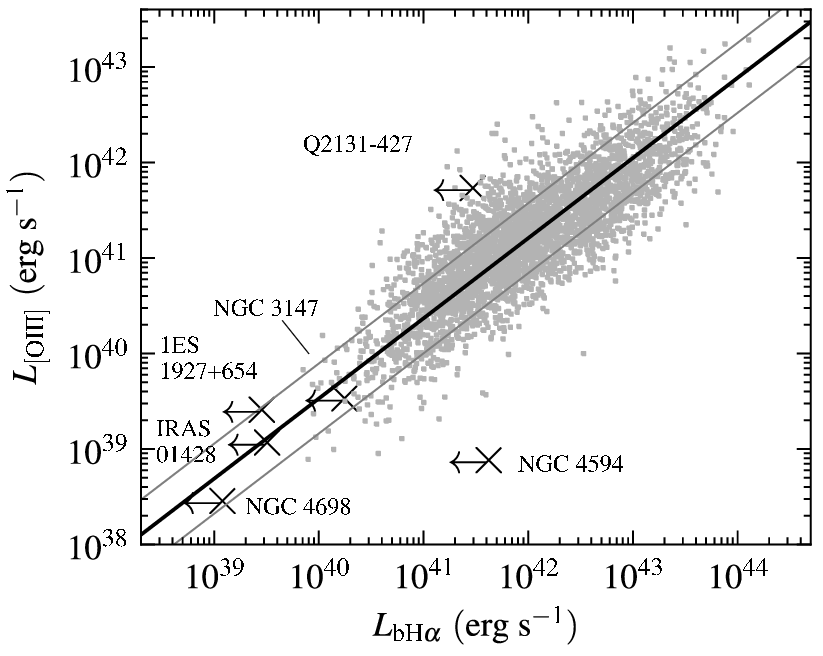}
\caption{The $\loiii$ vs. $\lbha$ relation as a probe for true type 2 AGN. The T1 objects are shown as gray dots. The best power law fit and dispersion (eq. 3) are indicated by black and gray lines. Also shown are upper limits on $\lbha$ in six true type 2 candidates from Panessa et al. (2009), Shi et al. (2010), and Tran et al. (2011). These objects have an apparently unobscured X-ray emission, but no broad \Ha\ detection. The upper limit on $\lbha$ depends on the expected $\dv$, which is derived from the $\mbh$ and $\lledd$ estimates for each object (eq. 7). Except Q2131-427, all candidates have upper limits which are consistent with their expected $\lbha$ based on $\loiii$. The Q2131-427 upper limit is 2.5$\sigma$ from the expected $\lbha$, and within the distribution spanned by the T1 sample. Thus, the absence of a broad \Ha\ in these six objects is not significant.
}
\end{figure}

\subsection{$\cfnlr$}

In this section we derive $\cfnlr$ by assuming that in the NLR most incident ionizing photons are absorbed by the gas, and not by the embedded dust. This assumption is relaxed in the following section. 
The low density of the NLR clouds ensures that an \Ha\ recombination photon is emitted roughly $1/2.2$ times per ionization event, independent of the NLR conditions. Hence, the \Ha\ emission rate equals the emission rate of the ionizing photons multiplied by $\cfnlr/2.2$. The rate of ionizing photons is $1.7   \times 10^{54} \ L_{\rm b\Ha;\ 42}\ {\rm s}^{-1}$, derived from $L_{\nu>\nu_{1200\AA}}\propto \nu^{-1.57}$ (Telfer et al. 2002) and $\nln(1200\AA)/\lbha=70$. The latter ratio is based on the mean observed $\nln(1450\AA)/\lbha$ (eq. 5 in Paper 1), adjusted to $\nln(1200\AA)$ using a local slope of --0.72 (Telfer et al.), and multiplied by a factor of two to account for the mean amount of UV extinction in the T1 sample (see \S3.7 in Paper I). 

The observed number of \Ha\ photons may differ from the intrinsic value due to extinction by dust. The extinction can be due to an external dust screen, or occur within the NLR clouds.
We neglect extinction from an external dust screen, since even if all the UV-extincting dust affects the NLR, than $\lnha$ is reduced merely by $\sim15\%$. Dust external to the NLR that does not affect the mean UV emission is unlikely, as the NLR is emitted from much larger scales. 
Within the NLR clouds, 50\% of the photons directed into the cloud are probably subject to high optical depth and converted to IR radiation. The other half of emitted photons are not expected to be extincted, as the dust opacity at optical frequencies is significantly lower than at ionizing frequencies (Groves et al. 2004a). 
Therefore, we assume that the intrinsic $\lnha$ is twice the observed $\lnha$. 

Equating the $\Ha$ emission rate to the emission rate of the ionizing photons, we get
\begin{equation}
\frac{2\ \lnha}{h\nu_\Ha} = 1.3 \times 10^{54}\ \frac{\lbol}{10^{44}\ \ergs}   \frac{\cfnlr}{2.2} ,
\end{equation}
where we used $\lbol/\lbha=130$ (Paper I). From eqs. 1 and 8, we get $\cfnlr=0.04$ at $\lbol=10^{45.5}\ \ergs$ and $\cfnlr=0.4$ at $\lbol=10^{42.5}\ \ergs$.

\subsection{Why does $\cfnlr$ decrease with $\lbol$?}

Why does $\cfnlr$ drop with luminosity in AGN?
A straightforward option is that the solid angle subtended by the gas surrounding the nucleus decreases with increasing $\lbol$.
Alternatively, the fraction of ionizing photons absorbed by dust embedded in the circumnuclear gas may increase with $\lbol$, thus decreasing the measured $\cfnlr$, which was derived assuming all ionizing photons are absorbed by the gas. In this section, we quantify the ratio of dust to gas absorption of a given cloud, and then discuss how to differentiate between these two possible drivers of the $\cfnlr$ vs. $\lbol$ trend.

\subsubsection{Dust bounded clouds}

For a given density, the gas opacity is roughly proportional to $U^{-1}$, where $U$ is the ionization parameter, compared to the dust opacity which is independent of $U$. Therefore the ionization rate, and consequently the recombination line flux, will increase with $U$ only up to a threshold $U_0 \sim 0.01$, where the dust and gas opacities are comparable (Laor \& Draine 1993). At $U>>U_0$, dust dominates the opacity, only $\sim U_0 / U$ of the ionizing photons are absorbed by the gas, and the emitted recombination line flux is only weakly dependent on $U$. This local effect persists when integrating over the entire cloud, using the $U$ at the cloud surface\footnote{Defined by neglecting the density gradient of the ionized zone.}
(Netzer \& Laor 1993), and also when accounting for the effect of radiation pressure on the density profile (Dopita et al. 2002, fig. 8 there). Hence, $U > U_0$ clouds are `dust bounded'. 

Assuming a higher $\lbol$ implies a higher mean $U$, and that a significant number of NLR clouds are dust bounded, then at higher $\lbol$ a larger fraction of ionizing photons are absorbed by dust, and the measured $\cfnlr$ will decrease. Evidence for the existence of these clouds was presented by Netzer \& Laor (1993), which showed that dust bounded clouds naturally explain the spatial gap between the BLR and NLR, and why the intermediate region emits mainly in the IR (the `torus'). Also, dust bounded clouds can account for the small scatter in $U$ observed in the NLR (Dopita et al. 2002, Groves et al. 2004b).

\subsubsection{Is the $\cfnlr$ vs. $\lbol$ trend due to dust bounded clouds or a drop in $\Omega$ with $\lbol$?}

We denote the covering factor of the circumnuclear gas, which can be different from $\cfnlr$ due to dust absorption, as $\Omega$. How do we differentiate between an $\Omega$ independent of $\lbol$, coupled with dust bounded clouds, and a decreasing $\Omega$ with $\lbol$? 
As mentioned in \S6.3, $\sim 50\%$ of the emission line photons are inward bound into the cloud, and will probably be converted to IR emission. Therefore, the fraction of ionizing photons initially absorbed by \hi, determined by $U$, will have a weak effect on the final amount of IR energy emitted from a cloud. Specifically, the IR emission is independent of $U$ up to a factor of $\sim 2$, in contrast with the strong effect of $U$ on the emission lines described above. Hence, by comparing $L_{\rm IR}/\lbol$ with $\lbol$ one can measure $\Omega$ directly. If dust bounded clouds dominate the trend seen in Fig. 3, then $L_{\rm IR}/\lbol$ should be roughly constant with $\lbol$, while if a decrease in $\Omega$ creates the trend, then we expect $L_{\rm IR}/\lbol \propto \lbol^{-0.3}$. 

A trend of decreasing $L_{\rm IR}/\lbol$ with $\lbol$ has been measured by Maiolino et al. (2007), Gallagher et al. (2007) and Treister et al. (2008). Maiolino et al. compared $L(6.7\mic)$ with $L(5100\AA)$ in 50 type 1 AGN with $44 < \log \lbol < 48.5$, and found $\Omega \propto \lbol^{-0.18}$. 
Gallagher et al. compared $L(1\mic - 100\mic)$ with $L(0.1\mic - 1\mic)$ in 234 quasars, and found a decrease by a factor of $\sim 2$ in $L_{\rm IR}/L_{\rm opt}$ over the luminosity range of $\log\ \lbol = 44.5 - 47$ (see their fig. 2). However, Gallagher et al. interpret this weak trend as due to increased reddening of the optical at low luminosity, and not as a change in $\Omega$. Treister et al. compared $L(12\mic)$ with the sum of optical and UV emission in 230 AGN, and found $\Omega \propto \lbol^{-0.14}$ at $44 < \log \lbol < 47.5$. 
The $L_{\rm IR} / \lbol \propto \lbol^{\sim -0.15}$ found in these studies is flatter than the $\lnha/\lbol \propto \lbol^{-0.3}$ found here, and the $\lbol$ measured are higher than the range of $\lbol$ spanned by the T1 sample. If the trend in $\Omega$ implied by these studies indeed continues to the lower $\lbol$ spanned by the T1 sample, than there seems to be a combined effect, both a decrease in $\Omega$ with $\lbol$, and a significant amount of dust bounded clouds. In a following paper, we compare $L_{\rm IR}$ with $\lbol$ in the \smpsz\ objects of the T1 sample.

\subsubsection{The type 1 / type 2 ratio}

The value of $\Omega$ decides the long standing question of the unobscured AGN / obscured AGN ratio, and its dependence on AGN characteristics. 
We note that the results of a type 1 / type 2 ratio study based on a 2--10 \kev\ selected AGN sample, are consistent with the IR studies mentioned in the previous section, at the same luminosity range (Hasinger 2008). Though, these results have been disputed by Reyes et al. (2008) and Hopkins et al. (2009). Furthermore, the increase of $L_{\rm NLR}/L_{\rm BLR}$ with decreasing $\lbol$ implies a stronger host dilution of the broad lines compared to the dilution of the narrow lines at low luminosity, which may cause misclassification of an unobscured AGN as an obscured AGN (see \S6.1 and \S6.2). If this effect is not accounted for, the fraction of type 2 AGN will seem to increase with decreasing $\lbol$, even if the true obscured fraction remains constant.

\section{Conclusions}

We analyze the dependence of the narrow \Ha\ and $\oiii\ \lambda5007$ luminosities on AGN properties in \smpsz\ $z<0.3$ type 1 AGN. We find the following:
\begin{enumerate}
\item The mean $\loiii$ and mean $\lnha \propto \lbha^{0.7}$, for $10^{40} < \lbha < 10^{44.5}\ \ergs$. 
\item Using the Paper I result that $\lbha \propto \lbol$, this trend implies a decrease in relative NLR luminosity with $\lbol$.
\item The key AGN characteristic driving the relative decrease in NLR luminosity is $\lbol$, rather than  $\mbh$ or $\lledd$.
\item The simple power law dependence of the relative NLR luminosity on $\lbol$ breaks at $\mbh \gtrsim 10^{8.5}\ \msun$, probably due to the dominance of radio loud objects, and at $\lledd \lesssim 10^{-2.6}$, probably due to the transition to LINERs.
\item The most likely mechanism behind this trend is a decrease in $\cfnlr$ with increasing $\lbol$. We derive $\cfnlr=0.4$ at $\lbol = 10^{42.5}\ \ergs$, and $\cfnlr=0.04$ at $\lbol = 10^{45.5}\ \ergs$.
\end{enumerate}

The implications of the decrease in relative NLR luminosity are:

\begin{enumerate} 
\item Intermediate type AGN dominate the type 1 AGN population at $\lbol < 10^{44}\ \ergs$. 
\item Intermediate type AGN at $\lbol < 10^{44}\ \ergs$ are generally not partially absorbed AGN. 
\item The implied \oiii\ bolometric correction factor in type 1 AGN changes from 3\,000 at $\lbol=10^{46}\ \ergs$ to 300 at $\lbol=10^{42.5}\ \ergs$.
\item The upper limit on $\lbha$ in six true type 2 candidates is consistent with their expected $\lbha$ based on their measured $\loiii$. An unusually high $\lx/\loiii$ ratio in true type 2 AGN candidates may indicate dust absorption of their line emission, rather than a physical absence of the BLR.
\item The decrease of $\cfnlr$ with $\lbol$ may be due to a larger fraction of ionizing photons absorbed by dust within the gas, 
or by a decreasing covering factor of all circumnuclear gas.
\end{enumerate} 

We thank the expert and knowledgeable referee for suggestions that significantly improved the paper, and Brent Groves for illuminating us on the typical EW of stellar absorption features. This publication makes use of data products from the SDSS project, funded by the Alfred P. Sloan Foundation, data from GALEX supported by NASA, and from the ROSAT Data Archive of the Max-Planck-Institut für extraterrestrische Physik (MPE) at Garching, Germany.

\appendix
\appendixpage

\section{Host subtraction}

As described in \S2.2, we adjusted the Yip et al. (2004) first eigenspectrum (ES1), which is used to model the host galaxy. In the following wavelengths, we replace the flux densities of the Yip et al. ES1, with the appropriate flux densities from the Hao et al. (2005) ES1: 
4835\AA--4880\AA; 4985\AA--5020\AA; 6285\AA--6320\AA; 6540\AA--6600\AA; 6700\AA--6750\AA. 
We use the $0.06<z<0.12$ Hao et al. ES1, normalized by 0.014.

In Figure A1, we show an example of the fitting procedure near \Ha, on a host dominated spectrum.

\begin{figure}
\includegraphics{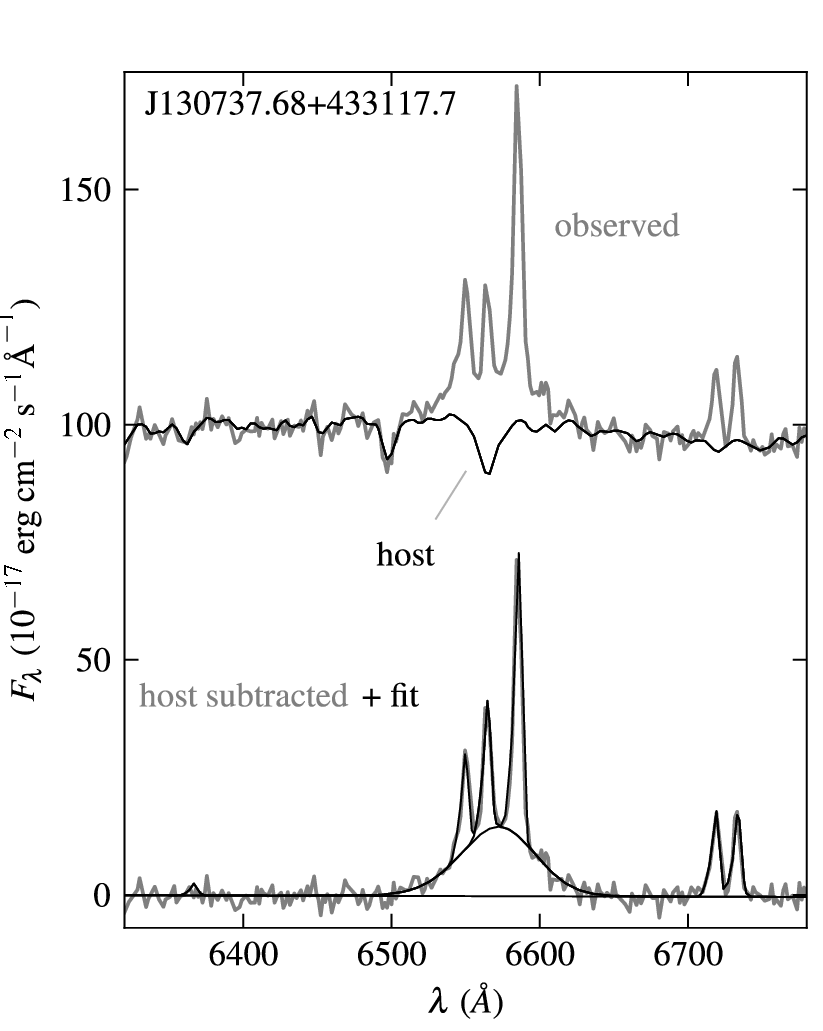}
\caption{An example of the fitting procedure near \Ha, on a host dominated spectrum. The plotted spectra are the observed SDSS spectrum, the fit host contribution, the host-subtracted spectrum, and the emission line fits. Note how the prominence of the narrow \Ha\ increases after subtracting the host. } 
\end{figure}

\section{Narrow line measurements}

The fits of the narrow \Ha\ and \oiii\ $\lambda 5007$ are described in \S2.3 -- \S2.4, and the measured fluxes are listed in Table 1. Here, we discuss the typical errors in the measurements, the derivations of upper limits when the emission lines are weak, and the sensitivity of the fluxes to the parameters of the algorithm.

\subsection{\oiii}

The \oiii\ is fit by a 4\th-order Gauss-Hermite (GH). In high S/N spectra, the profile may be too complex for the GH to achieve an acceptable fit. Therefore, we add to the formal error (derived from the spectrum errors published by SDSS) the difference between the data and the fit at $5007\pm10\AA$\footnote{The mean relative offset between the data and the fit is 2\%.}.  
In 99\% of the objects, the total error on $\foiii$ is $<30\%$ of $\foiii$. Out of the remaining 43 objects, in 32 the \oiii\ is not blended with other components, so we derive $\foiii$ by simply summing the continuum subtracted flux. In four objects only an upper limit on $\foiii$ can be deduced (see below). In three objects only the continuum was badly fit, so we use the measured $\foiii$. The remaining four objects are marked in Table 1 as having a large error in $\foiii$.

In 16 objects (0.5\% of the T1 sample) the fit \oiii\ flux density is $<3.5$ times the local flux density error. The tabulated $\foiii$ of these objects are upper limits, derived by assuming a Gaussian profile with this flux density and the width of the narrow \Ha.

\subsection{Narrow \Ha}

The error on $\fnha$ is calculated as described above for $\foiii$\footnote{The mean relative offset between the data and the fit is 4\%.}. In 97\% of the T1 sample, the relative error is $<30\%$. Of the 107 remaining objects, 
in 40 only an upper limit on $\fnha$ can be deduced. In three objects only the top of the broad \Ha\ was badly fit, so we use the measured $\fnha$. The remaining 64 objects (1.9\%) have poor fits, though due to their small number we do not attempt to further improve the algorithm. They are marked in Table 1.

Our algorithm can robustly detect the narrow \Ha\ if its flux density is above three times the local flux density error. The 73 objects with lower flux densities are marked as upper limits, derived by assuming a Gaussian profile with this flux density and the width fit to the other narrow emission lines.

Also, we note that the algorithm chooses the lowest $\chi^2$ from several fit attempts, which differ by the constraints on $\hc$ and $\hd$, or by the number of GH coefficients used to model the broad \Ha\ (\S2.4). We now quantify the effect on $\fnha$ of these different constraints. The $\hc$ and $\hd$ of the narrow lines are either set to zero, set to equal the $\oiii$ values, or are optimized during the fit. In 68\% of the objects the difference in the derived $\fnha$ between the three options is $<0.03$ dex, while in 90\% of the objects the difference is $<0.18$ dex. Choosing a different number of GH coefficients for the broad \Ha\ component changes $\fnha$ by $<0.02$ dex in 68\% of the objects and by $<0.07$ dex in 90\% of the objects.

\section{SF contribution to $\loiii$}

In this section we evaluate the mean SF contribution to $\loiii$, in half-decade $\lbha$ bins of the T1 sample. The mean host $\luv$ was derived in Paper I, by subtracting the expected AGN UV emission, based on $\lbha$, from the mean observed $\luv$. To convert $\luv$(host) to $\loiii$(host), we compare these two quantities in the 66 SDSS star forming galaxies described in \S3.3 of Paper I. These SF galaxies have complete UV detections, and span $10^{41.5}<\luv<10^{44}\ \ergs$ and $0.01 < z < 0.25$. We find
\begin{equation}
 \log\ L_{\rm \oiii;\ 43} = 0.77 \log\ L_{\rm UV;\ 43}  - 3.15
\end{equation}
where luminosities are given in units of $10^{43}\ \ergs$. The implied mean SF contribution to $\loiii$ is shown in Fig. 4. 

\label{lastpage}

\end{document}